# SOEN 6471 Team Project

## Project Milestone 1 – Team 9

Manpreet Kaur
Computer Science and Software Engineering Department
Concordia University
Montreal, Canada
manp_k@encs.concordia.ca

Sukhveer Kaur
Computer Science and Software Engineering Department
Concordia University
Montreal, Canada
suk_ka@encs.concordia.ca

Savpreet Kaur
Computer Science and Software Engineering Department
Concordia University
Montreal, Canada
sav_kaur@encs.concordia.ca

Aman Ohri
Computer Science and Software Engineering Department
Concordia University
Montreal, Canada
a_ohr@encs.concordia.ca

Ravjeet Singh
Computer Science and Software Engineering Department
Concordia University
Montreal, Canada
rav_si@encs.concordia.ca

Baljot Singh
Computer Science and Software Engineering Department
Concordia University
Montreal, Canada
bal_si@encs.concordia.ca

Navkaran Singh
Computer Science and Software Engineering Department
Concordia University
Montreal, Canada
navk_sin@encs.concordia.ca

Ravenna Sharma
Computer Science and Software Engineering Department
Concordia University
Montreal, Canada
rav_shar@encs.concordia.ca

***Abstract-*** ***Software architecture consists of series of decisions taken to give a structural solution that meets all the technical and operational requirements [24].***
***The paper involves code refactoring. Code refactoring is a process of changing the internal structure of the code without altering its external behavior. This paper focuses over open source systems experimental studies that are DMARF and GIPSY. We have gone through various research papers and analyzed their architectures. Refactoring improves understandability, maintainability, extensibility of the code. Code smells were identified through various tools such as JDeodorant, Logiscope, and CodePro.***
***Reverse engineering of DMARF and GIPSY were done for understanding the system. Tool used for this was Object Aid UML. For better understanding use cases, domain model, design class diagram are built.***

***Keywords:*** MARF, GIPSY, LOGICSCOPE, JDeodoant, McCabe, PWD, Multi-Tier Architecture, Run-Time system, Common Object Request Broker Architecture (CORBA)

## Background

This document focuses on the background study of two open source systems named as DMARF and GIPSY. DMARF is a Distributed extension of MARF based on pipelined distributed computing systems [8] and is used to provide services to clients that have low computational power[1]. GIPSY is an open source platform designed to explore the properties of intensional and imperative programming languages [8].

### *I. OSS Case Studies*

#### *1) DMARF*

##### *1.1. Introduction*

MARF (Modular Audio Recognition Framework) is an open source research platform written in Java language [2][6][8] that facilitates the modular framework which holds variety of pattern recognition (voice, sound, speech, text), NLP (Natural Language processing and signal processing algorithms [2][5][6][8]. Classic MARF is flexible and extensible framework that holds the capability of addition of new algorithms into the library for experimental use [8]. MARF has several applications and most revolve around its recognition pipelines, for example Text- Independent and Speaker Identification Applications. This document focuses on the latter application.

### *1.2. Architecture*

MARF's architecture shown in fig.1 is based on pipelined approach where the backbone consists of four pipeline stages, that facilitates data communication in a chained manner [2][6][8], and is almost similar for every framework including Distributed MARF.

The four pipelined stages mentioned in the architecture are:

1. Sample loading
2. Pre-processing
3. Feature extraction
4. Training/Classification.

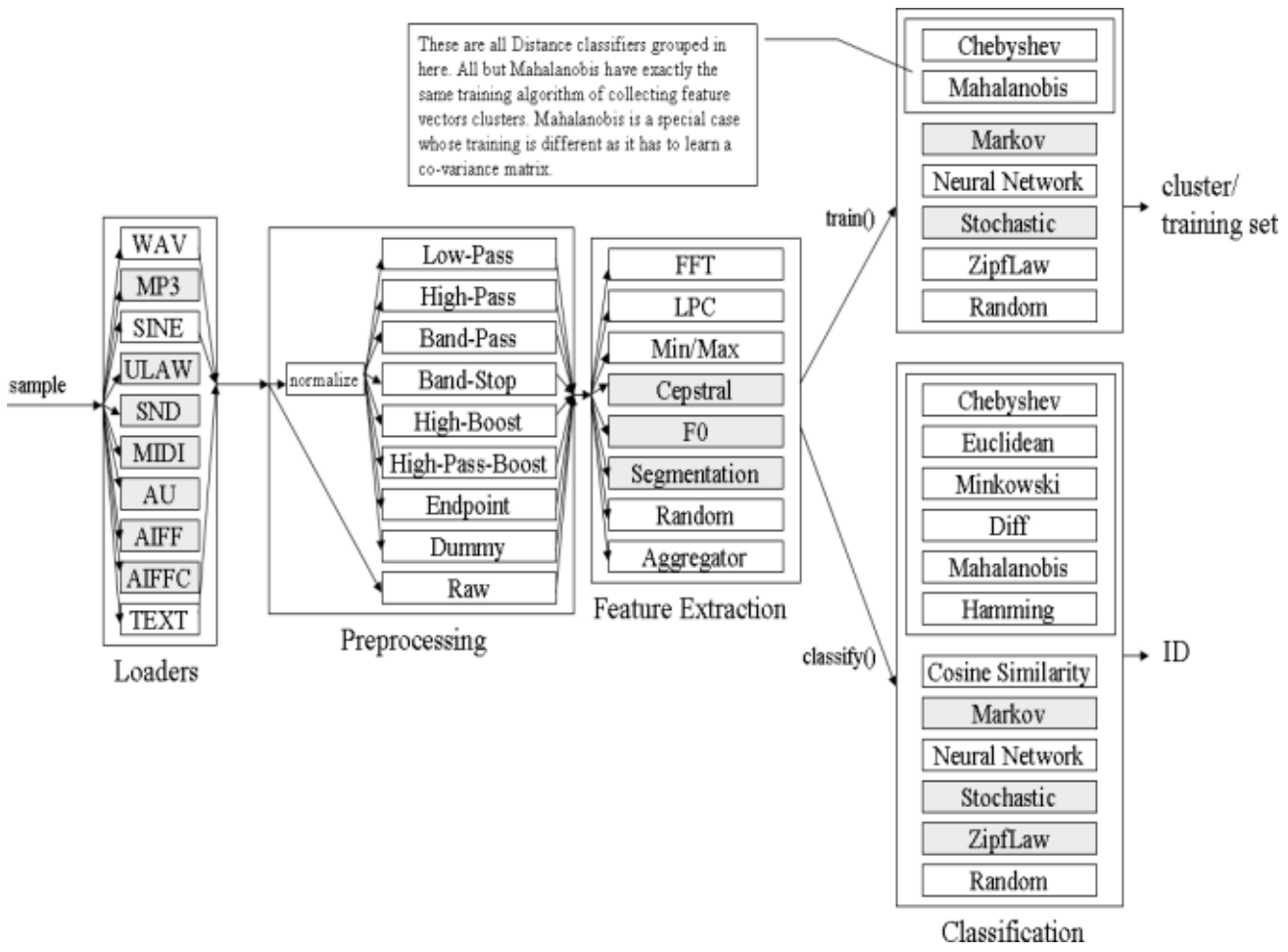

*Figure 1: Pipeline Data Flow [2]*

Algorithms (fig.1) in white boxes are implemented and which are in progress/stubs are expressed with gray boxes. SpeakerIdentApp is one of the prominent applications that test MARF's functionality [6][8].

These pipelined stages made to run distributive over the network using CORBA, XML-RPC web Services or Java RMI[2], runs stand-alone or act as library in applications[2][5][6][8]. This version is represented as **DMARF** (Distributed Modular Audio Recognition Framework), which further extended to allow node management using SNMPv3 protocol [8]. The DMARF architecture shown in fig.2 has a multi-level operational layers[6] where pipeline stages consists of front-end and back-end, having client application invocation, and grouping the similar kinds of algorithms like sample loading, pre-processing, feature extraction and training/classification[6].

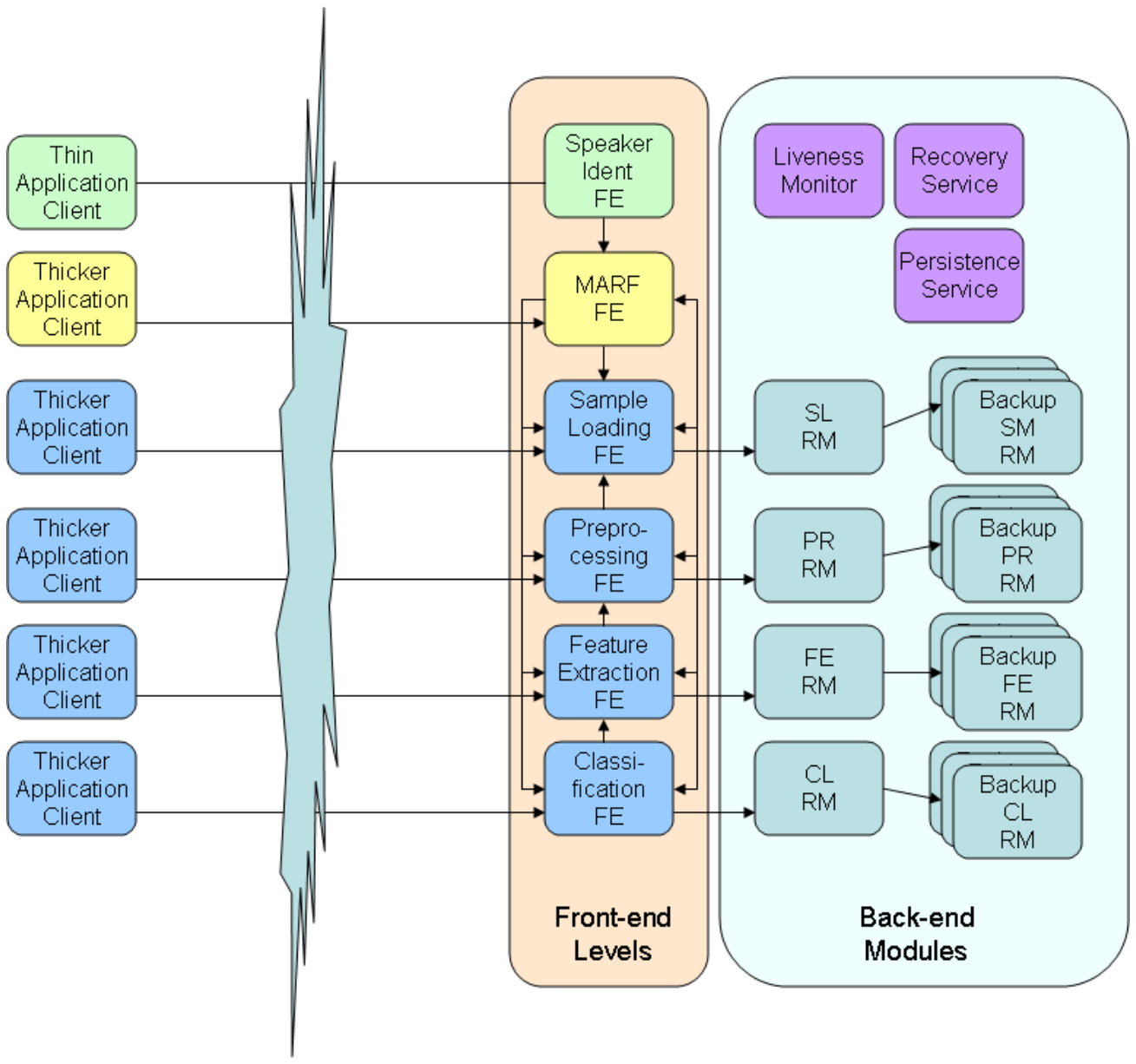

*Figure 2: The Distributed MARF Pipelines [1]*

Applications run on DMARF platform are fully architecture independent. It has two views as shown in fig.2:

**Module view**: DMARF applications are divided into two layers, one is the front-end which runs on the client side and server side and other is back-end that are services for the clients as shown in figure 2. The applications on the client side may be text-interactive or non-text interactive and applications which are on server side run on MARF pipelined approach. [1]

**Execution view**: It consists of runtime entities, communication paths and execution configuration. CORBA is networking technology, used for remote invocation. For remote method calls, RMI is used as base technology [1]. DMARF is a biologically inspired system employing pattern recognition, signal processing, and natural language processing helping us process audio, textual, or imagery data needed by a variety of scientific applications, e.g. biometric applications[6], included high volume of data processing for pattern recognition and biometric forensic analysis in a distributed manner and making web services more widely available over internet[5]is explained in further section of this document.

### *1.3. Principles and Services Offered*

The main principles of DMARF are [1]:

- Platform-independence
- Database-independent APIs
- Communication technology independence
- Reasonable efficiency
- Architectural consistency
- Simplicity
- Maintainability

Distributed MARF offers services listed below along with monitoring, backup recovery, disaster recovery modules[2][6]:

- Application Services
- General MARF Pipeline Services
- Sample Loading Services
- Pre-processing Services
- Feature Extraction Services
- Training and Classification Services.

### *1.4. Automatic DMARF (ADMARF)*

Autonomic Computing (AC) is self-managing characteristics of distributed resources to overcome the rapidly growing complexity of computing system management [3]. It helps to reduce the deployment and maintenance and increase the stability [3][4].

ASSL (Autonomic System Specification Language) has been used to formally specify and integrate special properties below into DMARF and code generation [4][6] An intrinsically complex system composed of multi-level operational layers[4]. A number of autonomic properties or basic elements of Automatic Computing specified for DMARF are mentioned below [3][4] and first two properties are explained in detail in further sections of the document:

1. Self-optimization
2. Self-protection
3. Self-healing
4. Self-configuring

**Self-Optimization**

ASSL approaches the problem of formal specification and code generation of autonomic systems (ASs) within a framework. ASSL provides a multi-tier specification model that is designed to be scalable and exposes a judicious selection and configuration of infrastructure elements and mechanisms needed by an AS [6].

The autonomic behaviour of DMARF is encoded in a special ASSL construct denoted as the SELF OPTIMIZING policy, which is specified at two levels [4]:

**AS-tier level**: At this tier, we specify a global system-level SELF OPTIMIZING policy and the actions and events supporting that policy. ASSL supports policy specifications with special constructs called Fluents and Mappings. Fluents are special states with conditional duration, while the mappings map actions to be executed when the system enters in such a state [4].

**AE-tier level**: It is the level of single AE, at this tier, we specify the SELF OPTIMIZING policy for the Classification stage nodes. A distinct AE is defined for every node. The specification has the partial specification of two AEs, each representing a single node of the Specification stage. At this level, self-optimization concentrates on adapting the single nodes to the most efficient communication protocol [4].

The algorithm behind the ASSL self-optimization model for DMARF is described by the following elements:

- Any time when ADMARF enters in the Classification stage, self-optimization behaviour takes place. [4]
- The Classification stage itself forces the stage nodes synchronize their latest cached results. Here each node is asked to get the results of the other nodes.[4]
- Before starting with the real computation, each stage, node strives to adapt to the most efficient currently available communication protocol. [4]

**Self-Protection**

DMARF cannot be used successfully in an environment where there are chances of malicious attacks and the protection of data is required. The protection of DMARF is less important in local networks but it becomes the key factor when it is supposed to run across the networks. Extending the functionality to include the security and protection functionality would be challenging because of the size and complexity of open source project like DMARF [3].

To implement the Self-protection in DMARF based systems, there should an authentic check on both source and target. This can be achieved by Specifying that each node in the pipeline needs to prove the identity to another. This can be achieved by issuing a proxy certificate to each node during the deployment and management phase [3].

Other benefits of Proxy certificate

- This proxy certificate can also be used in data privacy along public channels, especially when identities are required [3].
- Runtime protocol selection, it ensures availability, if the default communication becomes unavailable [3].

ASSL to specify the self-protecting behaviour in DMARF, the incoming messages must be secure in order to be able to process them.

Self-Protection Algorithm is shown in fig.3:

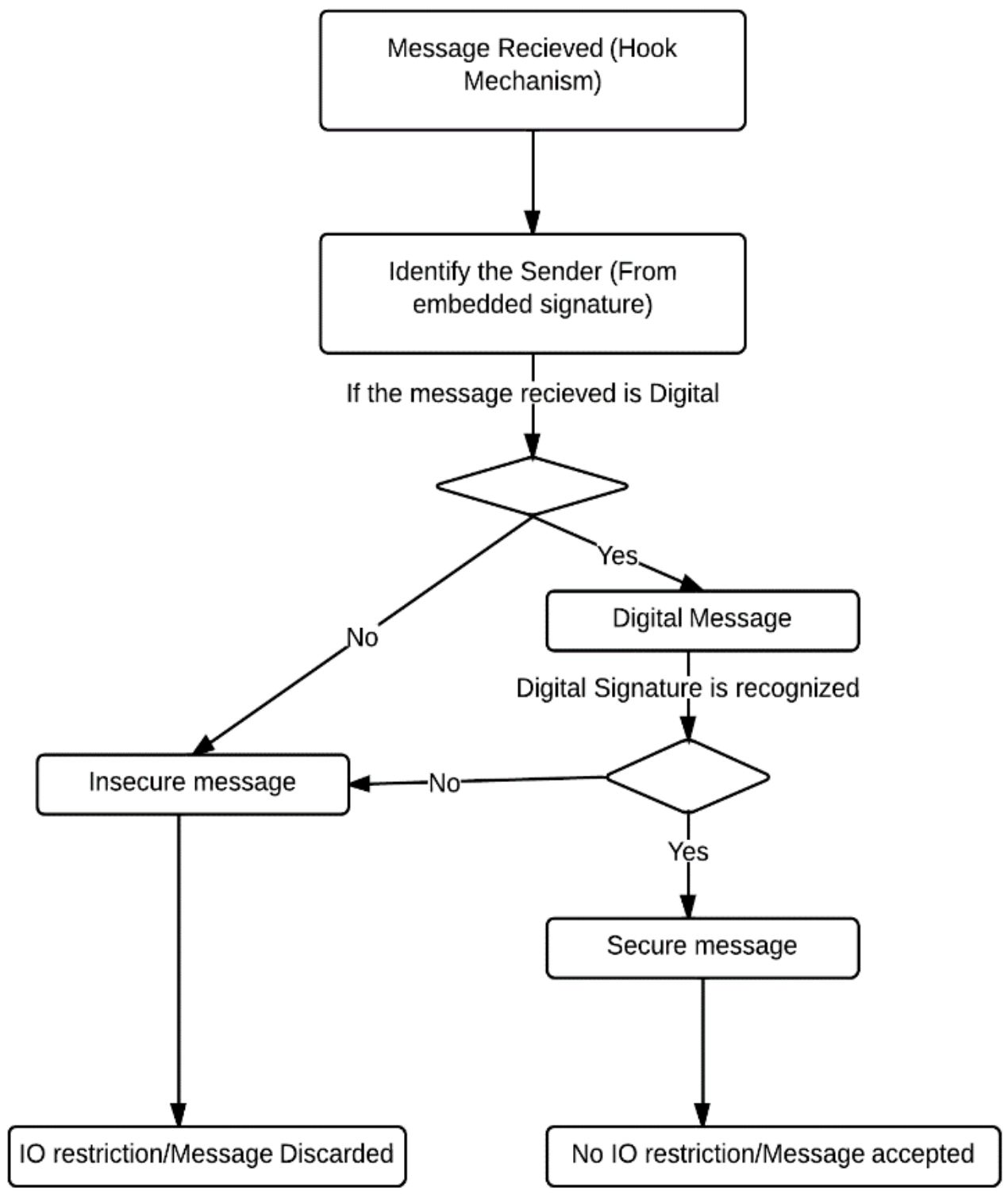

*Figure 3: Self Protection algorithm*

**IP tier Specification**

As per this specification, No entity can either send or receive a message that is not an ASSL-specified message. To implement this, two communication protocols at the ASIP tier for public message and at the AEIP tier for private message.

**AS Tier Specification**

In order to protect the AS from insecure public messages a self-management policy that handles the verification of any incoming public message is used.

**AE Tier Specification**

To deal with the privacy of private messages, a self-management policy identical to the policy specified at the AS tier is used. The ADMARF system in future will be able to fully function in autonomous environments[4][6], be those on the Internet, large multimedia processing farms, robotic spacecraft that do their own analysis, or simply even pattern-recognition research groups that can rely more on the availability of their systems that run for multiple days, unattended[6].

**Self-forensics**

The self-forensics have been introduced for the computer systems to automatically record their data, which can be used for computer crime investigations. Forensic Lucid is used as a forensic case specification language for automatic abstraction and event restoration of computer crime incidents. The language can state any event occurrences, time taken, their properties, as well as the context aware system model [7].

The self-forensics if applied on DMARF collects more forensics, data due to intrinsically more complicated relationship between modules as it is distributed and the configuration data related to configuration settings need to capture the configuration data related to the connection settings, protocols, and any other properties related to the distributed computing. Further, if there is similar configuration, then there could be a possibility to have multiple distributed/parallel training sets to be determined as well as multiple outputs can be produced on different nodes. During the life span of DMARF network of computing and nodes, there could be numerous pipelines. The pipelines then could be used for the analysis and interpretation of the cybercrime investigations and can be used for making data analysis decisions for incorrect data [7].

### *1.5. Towards Security Hardening of Scientific Demand-Driven and Pipelined Distributed Computing Systems*

Due to lack of security over the un-trusted local networks, various threats get associated with the computation involves maliciously induced incorrect computation results and cache poisoning hinders the availability and confidentiality in Demand-Driven (GIPSY) and Pipelined Distributed (DMARF) Computing Systems[8]. GIPSY is a modular framework mainly developed to investigate the lucid family of intensional programming languages. GIPSY basically deals with executable codes so it has more chances of encountering malicious code. Protocols like SSL or SSH, SNMPv3 are used to overcome some of the threats. Introduction of the GIPC framework in GIPSY, results in greater interoperability between intensional and imperative programming languages. GIPSY becomes truly distributed due to availability of DMS (Demand Migration System) but greater flexibility leads to security and demand monitoring related issues. DMF is introduced which is focused on the demand store with TAs and security was lies only with communication protocols but provide no mean over public unsecured network results in alteration of low level packets with corrupt/incorrect results[8]. DMARF extended with SNMP by implementing the proxy SNMPv2 agents and provides some security features for information management in v3 of the protocol with no data integrity and origin authentication assurance. Unlike GIPSY, DMARF does not hold any malicious code injection problem [8].

*Java Data Security Framework (JDSE)*

JDSF is a Java framework implemented for security researchers to lessen the security threats and evaluate various security algorithms and methodologies in a consistent environment and suitable for scope in research for GIPSY and DMARF. It aims at the data security aspects like Confidentiality, Integrity, Origin authentication, SQL randomization. It only focuses on data storage and cannot deal with DDoS and no provision for malicious code detection [8].

Security issues like confidentiality, Integrity, authentication and availability are also catered by JDSF in GIPSY and DMARF, using sub-framework and certification proxy approaches [8]. Confidentiality requirement is generally less applicable in GIPSY, where in DMARF it is applicable only in cyber-forensic case and has an export API to produce an SQL output. Availability is very difficult to ensure in a distributed environment and JDSF has no provision as well. Thus the proposed solution in the form of JDSF resolves most of the security issues, but unable to cater issues related to malicious code, thus other solutions are also taken into consideration to enhance security practices [8].

### 2) *GIPSY*

#### *1.1. Overview*

GIPSY (the General Intensional Programming System) is a continuing attempt to develop multi-language framework that aims to explore the domain of a lucid family of intensional programming of languages using demand-driven model [8]. GIPC is the compiler for GIPSY which is based on GIPL (Generic Intentional Programming Languages)[11].

#### *1.2. Need for GIPSY*

GIPSY was introduced by Dr. Joey Paquet and Dr. Peter Kropf at the turn of the millennia due to the aging and obsolescence of existing tools such as GLU (General Lucid) which was inefficient and inflexible. The General Intentional Programming System (GIPSY) gives a platform for the intensional programming exploration at long-term level.
The various other factors affecting were [8][10][11][12][14][16][18][19][15]:

- Language independent run time system was required for the execution of programs.
- Due to evolution of intensional programming, Lucid in particular, a system was required keeping in mind generality and adaptability in mind.
- Obeying the general architecture, a framework was required which can allow to replace the components.
- To make a system more efficient by invoking the optimization techniques.
- To begin a system which is different in the way that can develop a programming language this can build a bridge between them such as intensional and imperative paradigm.

#### *1.3. Goal*

GIPSY framework was designed to support intensional programming to achieve following goals [9][10][16][20]:

- To apply a multi-language development framework.
- To develop a flexible and scalable system which forms demand-driven and distributed multi-tier architecture consisting of loosely coupled components.
- To provide an interactive GUI to the user in the form of graphs which results in more flexible, usable, efficient and user-friendly experience.
- To incorporate peer to peer communication with propagation of Demands without having any information regarding the area of processing.

#### *1.4. Architecture*

Pre-Multi-Tier Era: The early architectural design of GEE had three categories as shown in fig.4 [15]

1. The Intensional Demand Propagator (IDP) generates demands.
2. The Intensional Value Warehouse (IVW) caches the values of computed demands
3. The Ripe Function Executor (RFE) does functional computation.

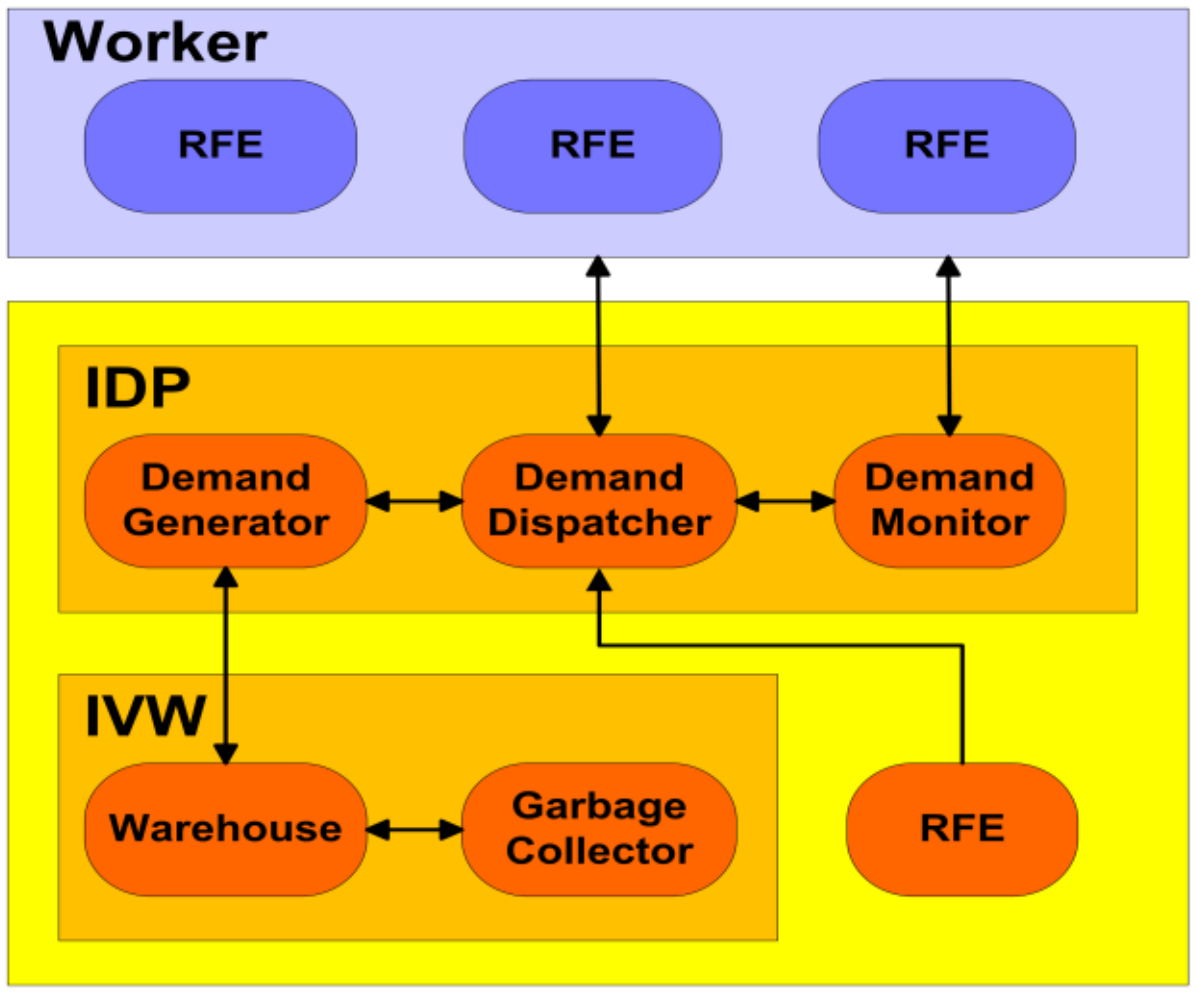

*Figure 4: Early GEE Architecture [15]*

Later, with research Demand Migration Framework (DMF) was introduced for the GIPSY runtime system and Demand Migration System (DMS), for migrating demands in heterogeneous and distributed environment. The purpose was to implement the DMF using Jini technology by Vassev and later JMS technology by Pouteymour. Figure 6 shows a GIPSY Demand Migration System.[15]

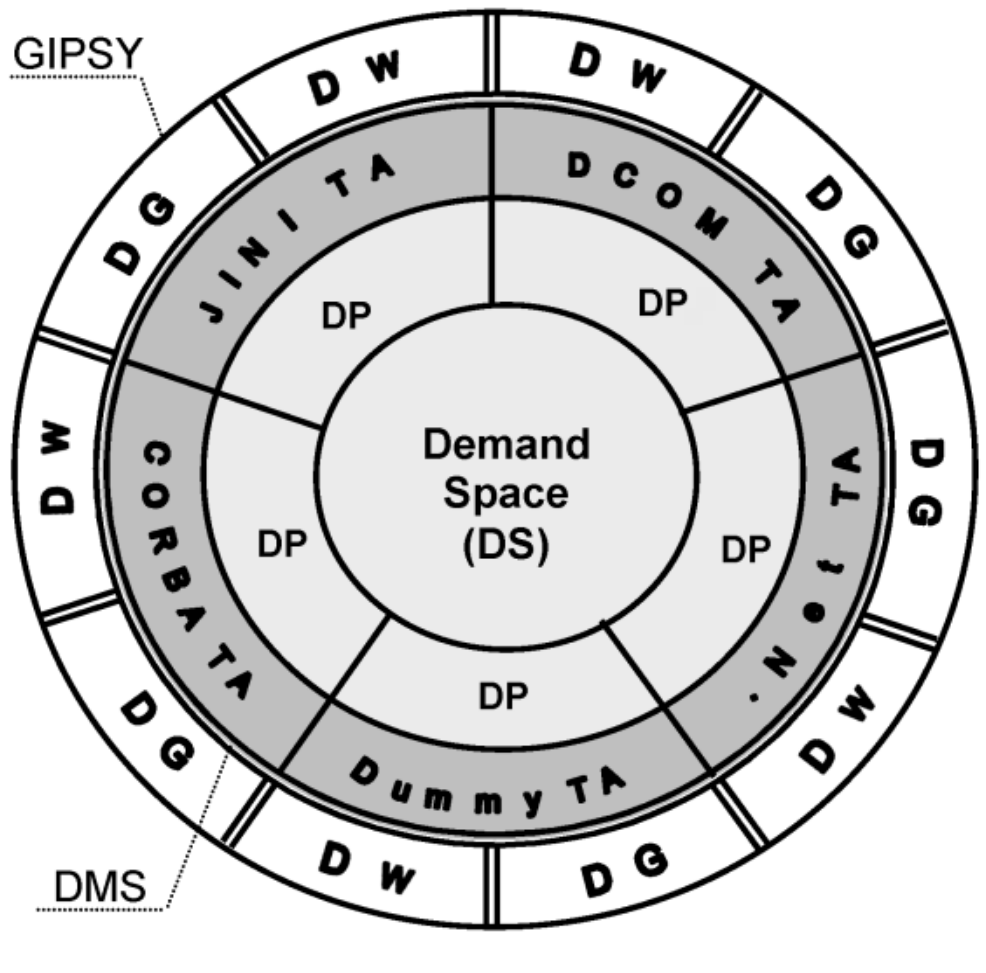

*Figure 5: GIPSY Demand Migration System [15]*

**The Multi-Tier Architecture**

With latest research multi-tier architecture come into existence for the GIPSY run time system. It contains some previous Demand Migration Framework features as well as four kinds of GIPSY tiers.[8][10][15]

- **Demand Store Tier (DST):** It acts as a communication system between other GIPSY tiers to migrate demands and provide the required demand storage.
- **Demand Generator Tier (DGT):** Its instance generates different types of demands by traversing the abstract syntax tree contained in a GEER.
- **Demand Worker Tier (DWT):** Process demands using procedure calls defined in GEERs and connects to DST to get back the pending demands and then hands over the computed demands to DST.

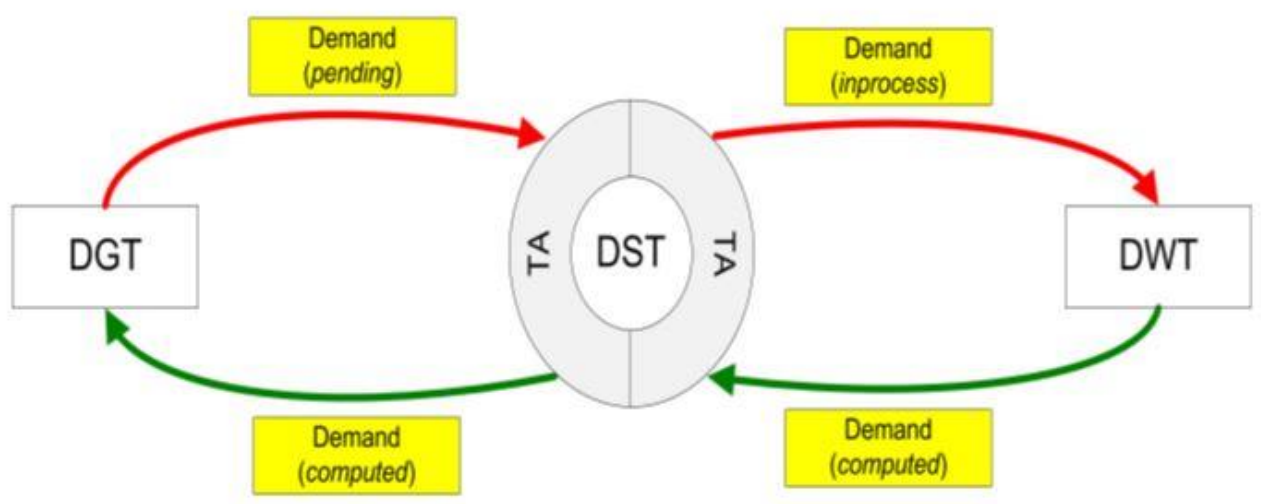

.

*Figure 6: The procedural demand migration among the DGT, the DST and the DWT[15]*

- **General Manager Tier (GMT):** It enables the registration of GIPSY node (GNs) to a GIPSY instance and then registration, allocation and de allocation of various GIPSY tiers.

Fig.7 indicates an example of three GIPSY instances analyzed by three distinguish colors running in six GIPSY nodes.

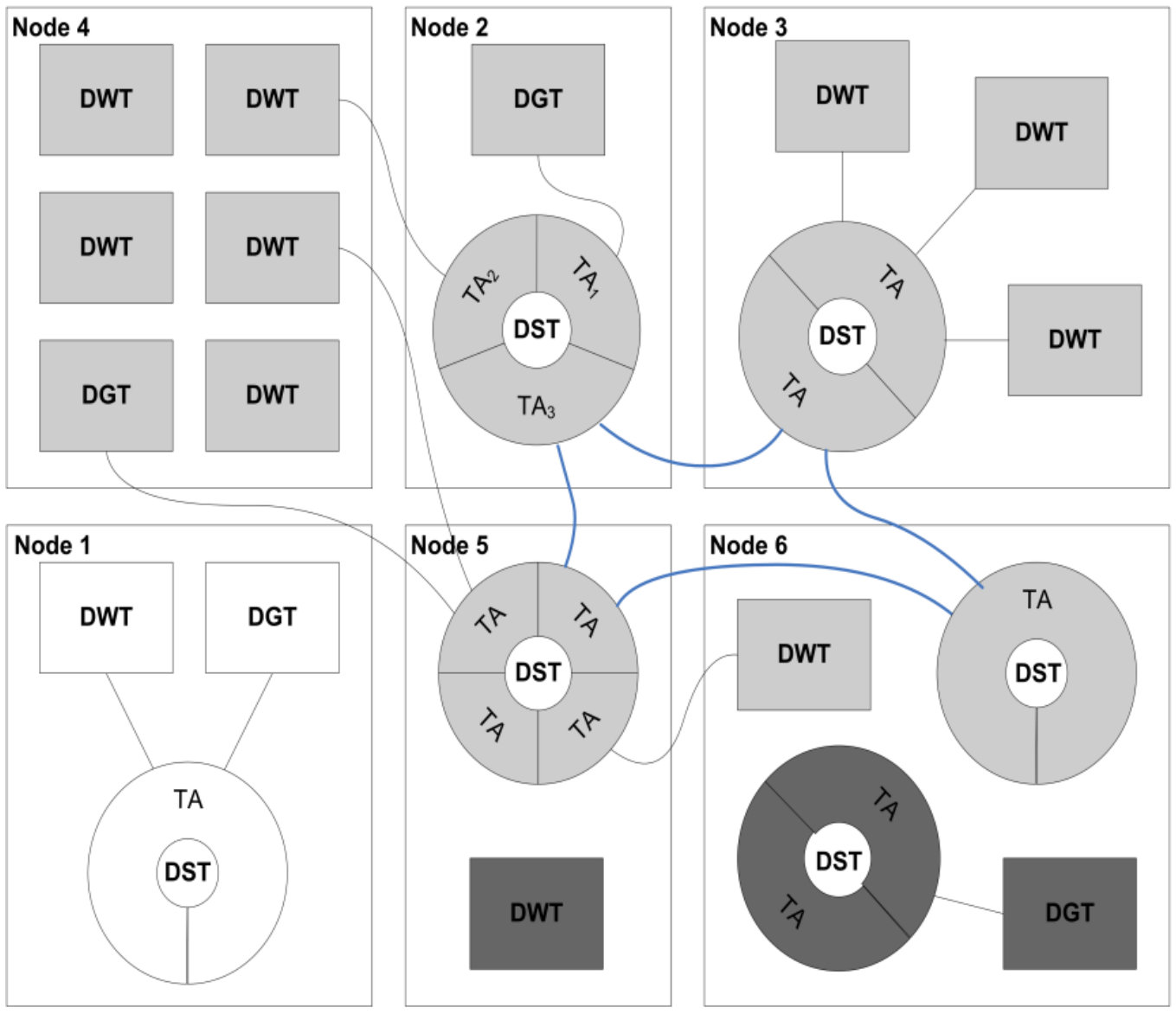

*Figure 7: Example of GIPSY instance [15]*

The Gipsy framework contains three main subsystems [8][10][15]:

1. **GIPC:** Flexible compiler called General Intensional Programming Compiler translates any intensional program into source language independent runtime resources.
2. **GEE:** A language Independent runtime system called General Education Engine. Runtime system comes to action by using that runtime resources provided by GIPC to execute the program in a demand driven and distributed manner.
3. **RIPE:** A component called Runtime Iterative Programming Environment. Provide a visual user interface, so the user can interact with the runtime system. The RIPE allows the dynamic interaction of the user and it permits to change the communication protocol. Moreover, there is a graphical representation of lucid programs in the form of data flow diagrams and collection of garbage values. The Compiler first translate Gipsy program into a source language independent General Intensional Programming Language**(GIPL)** program, then generate source language independent Generic Eduction Engine Resources **(GEER)** [10][15].

### *1.5. Autonomic GIPSY (AGIPSY)*

In order to reduce the workload on a complex GIPSY system, AGIPSY is designed to perform the self-configuration, self-optimization, self-healing, self-protection & self-monitoring on a complex multi-tiered distributed heterogeneous workloads system.[14]

The foundation of AGISPY lies in the ASSL (Automatic System specification language) framework.

Three major tiers of ASSL are:

1. AS Tier
2. AS Interaction protocol
3. AE Tier

Implementation of ASSL framework on AGISPY:

**AS Tier:** It specifies the service level objective of the AS, self-management policies, metrics and the architecture, Actions, events.

**AS Interaction protocol tier:** The ASSL Framework specifies an AS level Interaction protocol (ASIP) through the Public AS messages & Negotiation Protocol, Public communication channels & Public communication functions.

**AE Tier:** The ASSL Framework describes the individual AEs of the AS including the AE service level objective, AE self-management policies, friends, AE interaction protocol, recovery protocol, behaviour, outcomes, actions, events and metrics.

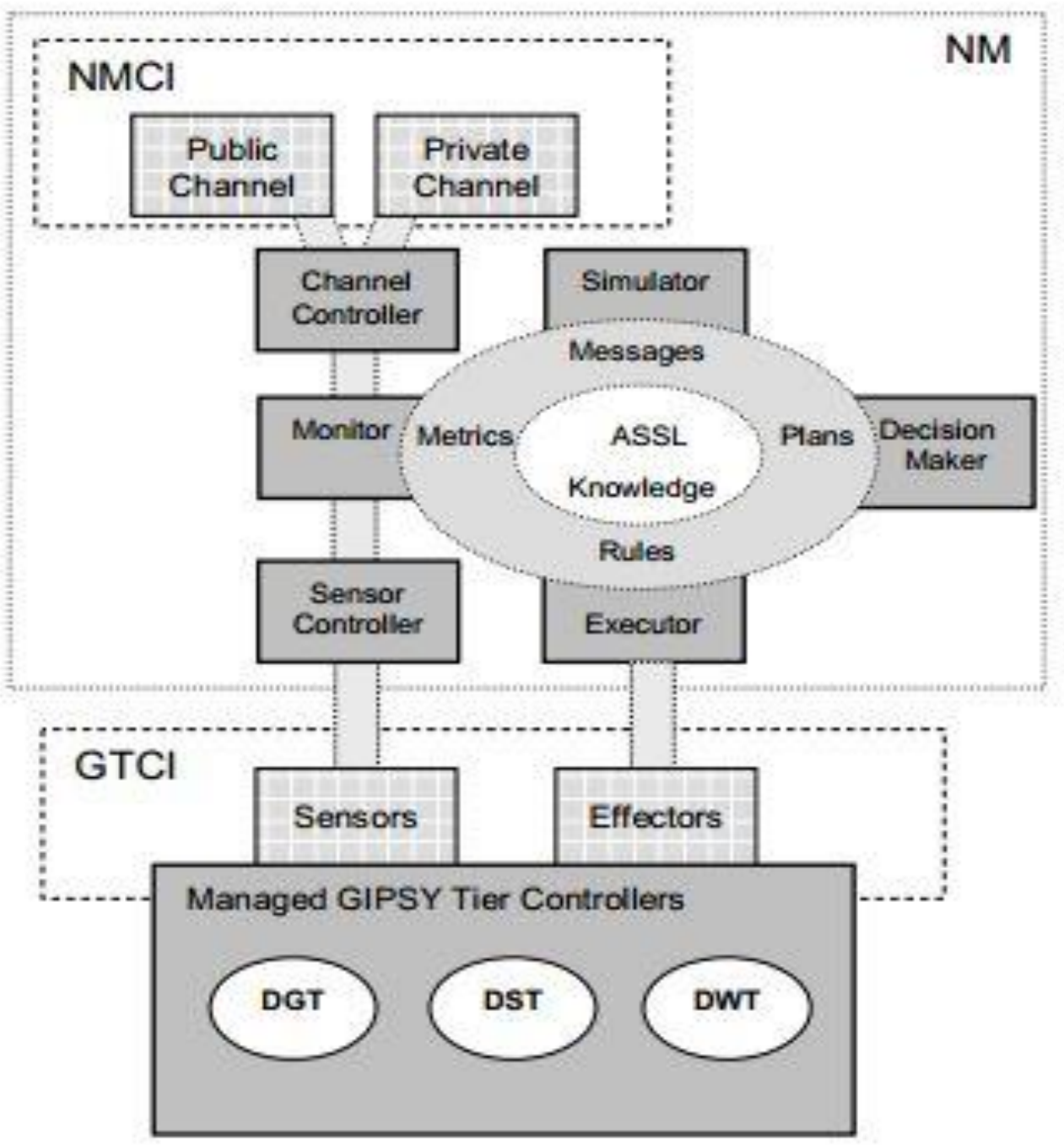

*Figure 8: GIPSY AE Architecture [14]*

Autonomic System Specification Language (ASSL) is a framework that specifies and generates an autonomic system. Forensic Lucid is an intensional context-oriented forensic case identification, modeling, and assessment language. It could be used in automatic recognition and reconstruction of event in digital forensic and investigation of incidents etc.

Need of the AGIPSY on a complex multi-tiered distributed heterogeneous system:

- Self-configuration
- Self-healing
- Self-optimization
- Self-protection

Like CHOP properties, the notion of SELF_FORENSICS policy for AS tier and AE is added. The addition of syntax and semantic support for the lexical analyzer, parser, and semantic checker and the addition of the code generator for JOOIP and Forensic Lucid to translating forensic events are the two major parts of the property introduction. The managed element specification of AE is used to encode module or sub-system to increase the forensic log depending on the criticality of the faults.[13]

*3) Summary of Case Studies*

DMARF is an open-source research platform and MARF's distributed extension that uses distributed pipeline stages and multi-level operational layer approach to communicate with each other to get the data they need in a chained manner [6]. DMARF cannot be used in an autonomous environment due to lack of design provisions and security, which leads to the requirement of self- adapting nature, such as self-optimization and self-protection [4][6]. ASSL approaches the problem of formal specification and code generation of autonomic systems (ASs) within a framework [6]. Thus, several principles of Automatic Computing have been applied to solve specific problems like security, performance, etc.[6]

Due to the aging of the conventional GLU, GIPSY was introduced to provide a more reliable, efficient and adaptable system and its advanced framework directed towards analyzing intensional programming language portrays compiled programs called GEER [8][10][13][15]. The inclusion of loosely coupled components that led to high flexibility, high scalability, increased usability and an organised system in demand driven, multi-language development framework and distributed multi-tier architecture was attained using the intensional programming that used GIPSY. It has its roots strengthened by the introduction of new proposed technologies such as AGIPS which would make it an intelligent system to perform all its major functionalities on its own such as self-healing, self-optimization, fault-tolerance etc[14]. Overall, GIPSY once fully implemented would be distinctly reliable that would even streamline the distributed execution of hybrid intensional necessary programs using JAVA.

TABLE I. Lists the readings that inform the background of this paper, and identifies the team member that was responsible for each.

TABLE I. CASE STUDY DISTRIBUTION AMONG TEAM MEMBERS

| Reference | Read by |
|---|---|
| Distributed Modular Audio Recognition Framework (DMARF) and its applications over web services.[5] | Raveena Sharma |
| On design and implementation of distributed modular audio recognition framework: Requirements and specification, design document[1] | Aman Ohri |
| Managing distributed MARF with SNMP[2] | Baljot Singh |
| Autonomic specification of self-protection for Distributed MARF with ASSL[3] | Manpreet Kaur |
| Towards autonomic specification of Distributed MARF with ASSL: Self-healing[4] | Navkaran Singh |
| Distributed Modular Audio Recognition Framework (DMARF) and its applications over web services.[6] | Ravjeet Singh |
| Towards security hardening of scientific distributed demand-driven and pipelined computing systems[7] | Sukhveer Kaur |
| Self-forensics through case studies of small to medium software systems[8] | Savpreet Kaur |
| Towards autonomic GIPSY[14] | Manpreet Kaur |
| The GIPSY architecture[10] | Savpreet Kaur |
| Unifying and refactoring DMF to support concurrent Jini and JMS DMS in GIPSY[12] | Sukhveer Kaur |
| Distributed educative execution of hybrid intensional programs. [16] | Navkaran Singh |
| Scalability evaluation of the GIPSY runtime system[15] | Baljot Singh |
| An interactive graph-based automation assistant: A case study to manage the GIPSY's distributed multi-tier run-time system.[9] | Ravjeet Singh |

| Reference | Read by |
|---|---|
| Towards a self-forensics property in the ASSL toolset. In Proceedings of the Third C* Conference on Computer. Science and Software Engineering[13] | Raveena Sharma |
| Advances in the design and implementation of a multi-tier architecture in the GIPSY environment with Java.[11] | Aman Ohri |

### 4) *METRICS*

We provide measurements of the following criteria for DMARF and GIPSY by using SonarQube.

**SonarQube** is an open source platform for Continuous Inspection of code quality. It offers analysis on duplicated code, coding standards, unit tests, code coverage, complex code, potential bugs, comments and design and architecture. [17]

We accomplished this using the following process:

- Ran the sonar runner (runner.bat) and make sure that server running on local host at port 9000.
- Copy the *sonar.project.propertiesfileandsoanr* folder file under both the project folder and edited its properties as the project folder location.
- Run Sonar-runner.bat.

The results of measurements are presented in Table below:

TABLE II. RESULTS OF DMARF AND GIPSY AFTER MEASUREMENT

| Measurement | DMARF | GIPSY |
|---|---|---|
| Java files | 1024 | 602 |
| Classes | 1054 | 666 |
| Methods | 7152 | 6262 |
| Lines of Java code | 77297 | 104073 |

Fig.9 and 10 shows the measurement results for GIPSY and DMARF

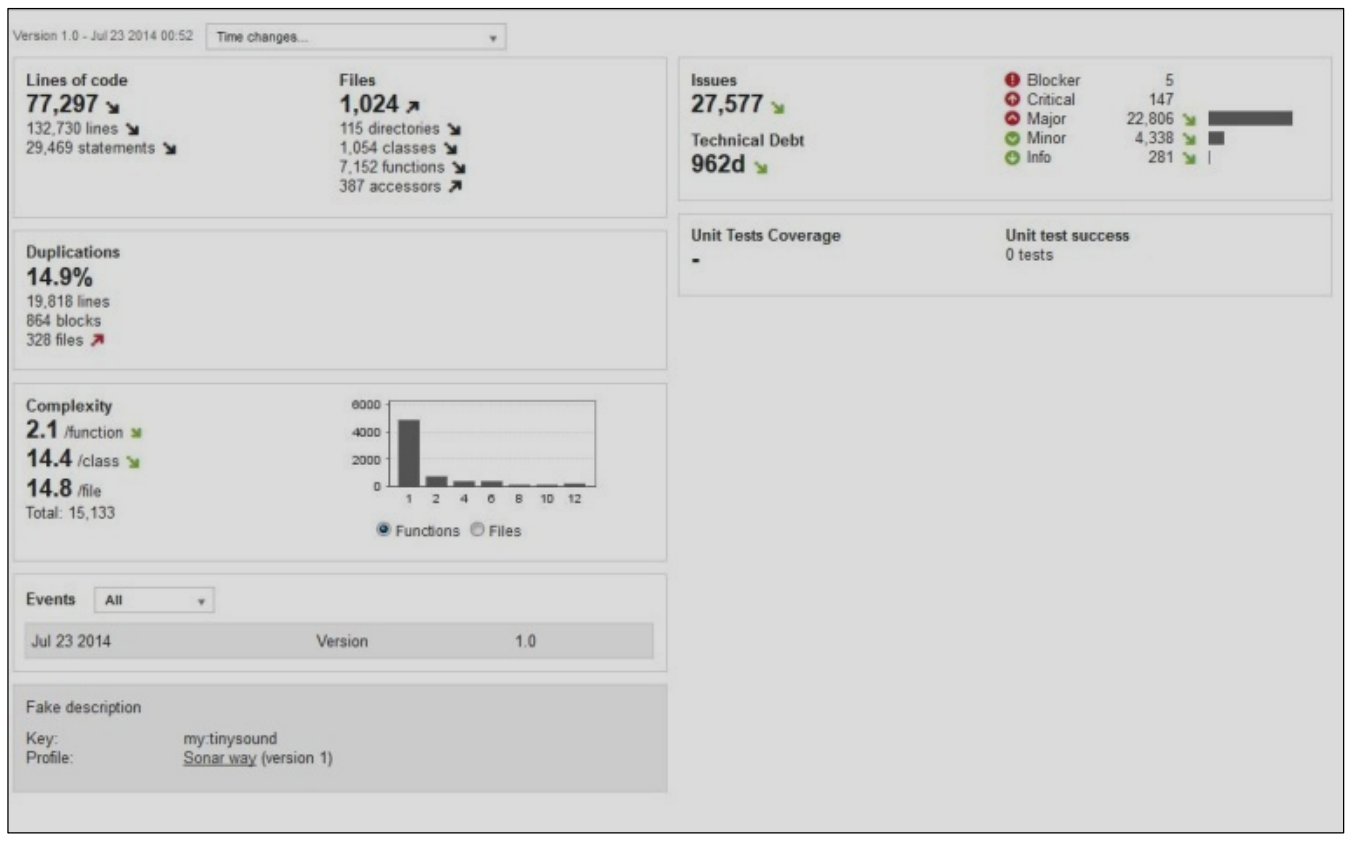

*Figure 9: Shows the measurement result for DMARF*

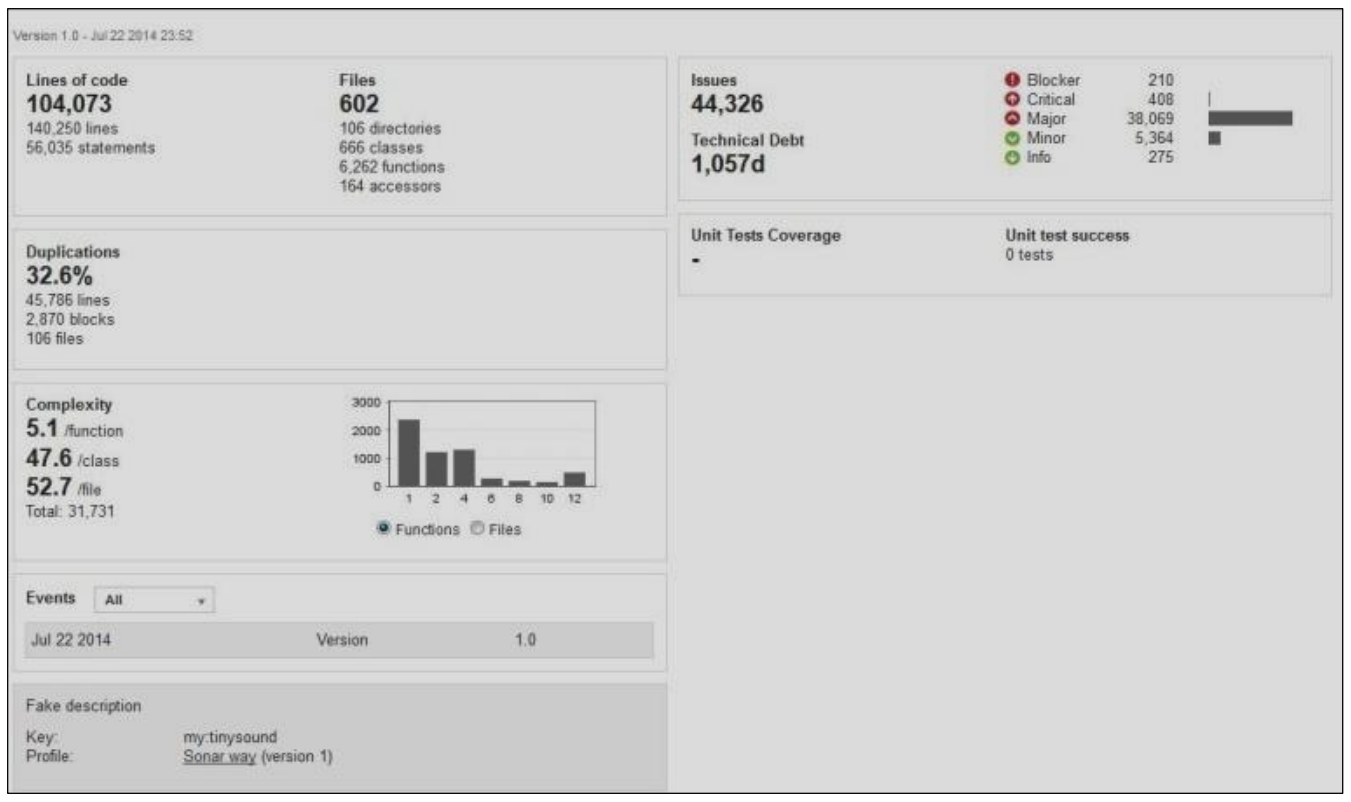

*Figure 10: Shows the measurement result for GIPSY*

## II. *Requirements and Design Specifications*

DMARF is designed for the pattern recognition and has several applications, which revolve around its recognition pipelines. The main purpose of DMARF is to make the pipeline distributed while maintaining the functioning of the traditional MARF.

GIPSY focuses long-term investigation of intensional programming. It uses a multi-tier architecture. GIPSY also supports Demand Migration Framework. GIPSY multi-tier runtime aims to research scalability.[15]

### 1) *Personas, Actors, and Stakeholders*

*a) DMARF*

*User Persona*

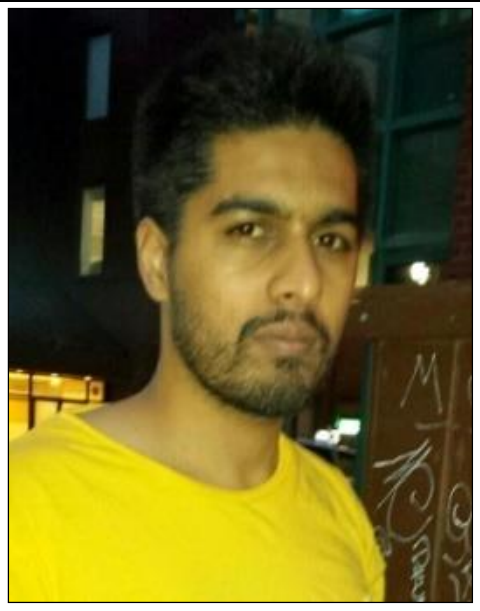

Persona: Master's student
Name: Ravjeet Singh

Ravjeet is a 26 years old master's student at the Concordia University, Montreal. Ravjeet has been doing his master's in the field of Software. He characterizes his persona as "Goal-Directed" and someone, who like the things to be easier to understand by others. Since childhood, Ravjeet had a great interest in computers, which urged him to take it as his major in studies. Apart from his studies, Ravjeet likes to socialize and, an active user in the field of learning to know the newly released technology.

As a part of his coursework, Ravjeet is doing a project in which he wants to use speaker identification for the purpose of making a machine run. The machine should be able to identify a certain frequency of voice/speech and, only take commands from a certain set of users. Then, the machine will be able to encrypt the message into its own language and be able to work on the commands provided. He feels using speaker identification in his project will help him in increasing security and, reliability of his system. Also, he feels by doing this the system can be operated from a remote site.

*Actor*

TABLE III. LIST OF ACTORS AND THEIR DESCRIPTION

| Actors | Description | Type |
|---|---|---|
| Users (Students) | A user provides the data sample in the form of audio(mp3) to the SpeakerIdentApp for audio recognition | Primary |
| Secondary servers | Secondary server holds the image of the main MARF server and update status at every transaction, In case primary server is down, the secondary takes up the charge. | Secondary |
| Storage | A software application that interact with the framework to capture, analyze and store data. | Primary |

*Stakeholders*

TABLE IV. LIST OF STAKEHOLDERS AND THEIR DESCRIPTION

| Stakeholders | Description |
|---|---|
| Users (Students) | Interacts with the system and initiate the processing proving Audio sample. |
| Developers | Persons responsible for the system development, implementations and code improvement |
| Architects | Individual/Team responsible to make the high level architectural design and dictates software coding standards, tools and platforms. |
| Support Team | The team responsible for system maintenance and authentication. and technical support |

*b) GIPSY*

*User Persona*

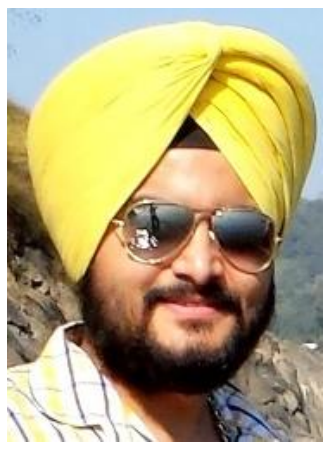

Gender: Male
Age: 24
Location: Montreal, QC
Internet usage: 5-6 hrs /day

Occupation: Student
Baljot is currently pursuing his Master's at Concordia University. Apart from studying he enjoys socialising with people. When he is free, he likes to go out with his friends for long walks and movies.
Baljot's aim is to excel in the field of software. He has great interest in doing research which provides a framework for a distributed multi-tier demand-driven evaluation of heterogeneous programs together with cyber forensic.
He has his focus over intensional logic to cyber forensic analysis that will aim at backtracking of events rebuilding so the evidence can be modeled by multidimensional hierarchal context and proofs are taken in an eductive manner of evaluation. This approach is improvement over finite state automata (FSA).
He has seen many changes in these last years in terms of the architecture of its research. The research is upgrading years by years and becoming faster and better.

*Actors*

TABLE V. LIST OF ACTORS AND THEIR DESCRIPTION

| Actors | Description | Type |
|---|---|---|
| Users (Students) | User inputs the data for the analysis and user register itself to form a node. | Primary |
| GIPSY Node | Multiple tiers are allocated inside the computer after registration is done. | Secondary |

*Stakeholders*

TABLE VI. LIST OF STAKEHOLDERS AND THEIR DESCRIPTION

| Stakeholders | Description |
|---|---|
| Users (Students/ Research analyst Teacher) | They provide the data for the system. |
| Developers | The developers generates the code for the functionalities and interactions. |

*2) Use Cases*

The Use case shows interaction between actor, stakeholders and system and provides an overview of all the requirements for a system in the form of essential model and also communicates the scope of development project. The fully dressed use cases for DMARF and GIPSY are mentioned in subsequent sections.

*a) DMARF*

TABLE VII. FULLY DRESSED USE CASE OF DMARF

| | |
|---|---|
| **User Case UC1** | To recognize the Audio Sample and provide results |
| **Primary Actor** | **Student**, **SpeakerIdent Application**, DMARF Platform, Storage set |
| **Secondary** | Secondary Server |
| **Actor** | |
| **Stakeholders and Interest** | **Student:** Analyse the audio sample sent to SpeakerIdent Application using DMARF platform and understands the architecture and working of DMARF for security purposes.<br>**Supporters:** Provides **the technical support to** Students if required. |
| **Preconditions** | SpeakerIdent Servant should be connected to MARF Servant to take the input data sample from the user.<br>MARF servant should have a sample **audio data** for processing. |
| **Post condition** | **MARF Servant** should provide result to the student on the bases of loaded data and should store the result in result set. |
| **Main success scenario** | 1. Student loads the audio data onto the **SpeakerIdent Servant**.<br>2. **MARF Servant** accesses the audio sample from SpeakerIdent servant and loads the sample data in Sample loader.<br>3. MARF Servant will invokes the **pipeline processing**.<br>4. A **Pre-Processer** that accepts the sample from sample loader and does the required initial processing of sample data.<br>5. A **Feature Extractor** accepts the pre-processed data and creates a set of required features, which will be useful for further processing and may invoke pre-processing for pre-processed sample.<br>6. Classifier invokes the feature extraction services for features and perform data **classification** and on the bases of analysis, it creates **Training set**.<br>7. **Result set** provides the result to MARF server on the bases of Training set created by the analyzer. |
| **Extension/ Alternatives scenario** | **If pipeline stage process crashes:**<br>It will use **message-logging** protocol so that a module could recover information concerning that module's data after a faulty processor has been repaired. |
| **Special requirements** | MARF Servant should be connected to the **storage set** |
| **Technology and data Variations list** | System should be designed in such so that any communication technologies, adapters or plug-ins can be added with little or no change to the main logic and code base. |
| **Open issue** | **DMARF** does a lot of writes (dumps) and long-running servers have a potential to have their **transaction IDs** be recycled after an overflow. |

*b) GIPSY*

TABLE VIII. FULLY DRESSED USE CASE OF GIPSY

| | |
|---|---|
| **Use Case UC2** | Intensional programming analysis system. |
| **Primary actor** | Student |
| **Stakeholders and interests** | Student: wants to analyse the intensional system.<br>Researcher: wants users to use the application in different fields.<br>Developer: wants to increase the usability of the application. |
| **Pre-condition** | Register to sample |
| **Success guarantee (Post-condition)** | Desired result is delivered |
| **Main success scenario (or basic flow)** | 1. Sample input (vulnerable code) is given to the system by the user through an interactive interface **RIPE** package interfacing **GMT.**<br>2. The program is then compiled in GIPSY compiler that is **GIPC**.<br>3. An intermediate source language independent program is generated, **GIPL.**<br>4. Through this **GEER** is generated that is also a source language independent program that contains lucid intensional and procedural identifier and **AST (Abstract Syntax Tree)** of GIPL program.<br>5. **DGT** traverses **AST** to fetch different types of demands. The pending demands in DST are retrieved by **DWT** and then the definition is searched. On finding the result DWT returns it to **DST**.<br>6. Then the **GEER** is passed to the **GEE**. The results that are in DST are executed using demand-driven eduction model.<br>7. Result comes out to be as more realistic evidence representation and more scalable and accessible to audience due to its simple nature. |
| **Extensions/ Alternative Scenarios** | If **DST** fails:<br>**DWT** has a local demand that stores computed demands in case DST fails. |
| **Special requirements** | Lucid programming should be installed. |

### 3) *Domain Model UML Diagrams*

Domain Model is a conceptual model describes the various entities, their attributes, roles, relationships and the constraints that governs the problem domain. We have designed the conceptual models for both the case studies DMARF and GIPSY in context to the fully dressed use cases explained in above sections. These models provides the concept view of how system works internally and with external actors.

#### a) ***DMARF*** *Conceptual Model*

In the Fig. 11 given below: The client interacts with system through SpeakerIdent client and sends audio sample data to SpeakerIdent Servant which transfer the sample to MARF which invoke four pipeline stages for further data processing. First pipeline stage-Sample loader loads the audio sample in the form of file and MARF Servant invokes the Pre-processor to process the sample file into preprocessed data which get invoke by Feature Extractor pipeline to extract the features from the samples and generate data vectors which act as input to Classification pipeline where MARF servant classify the extracted data vectors into transaction IDs. Logger maintains the message log and hold the details regarding transaction IDs and uses message-logging protocol so that a module could recover information after a faulty processor has been repaired.

The DMARF servant gets the result from database on the bases of instruction set and training set. In case the primary server is down due to some catastrophic or technical conditions, secondary server will takes up the charge and whole of the load will be handled by secondary server.

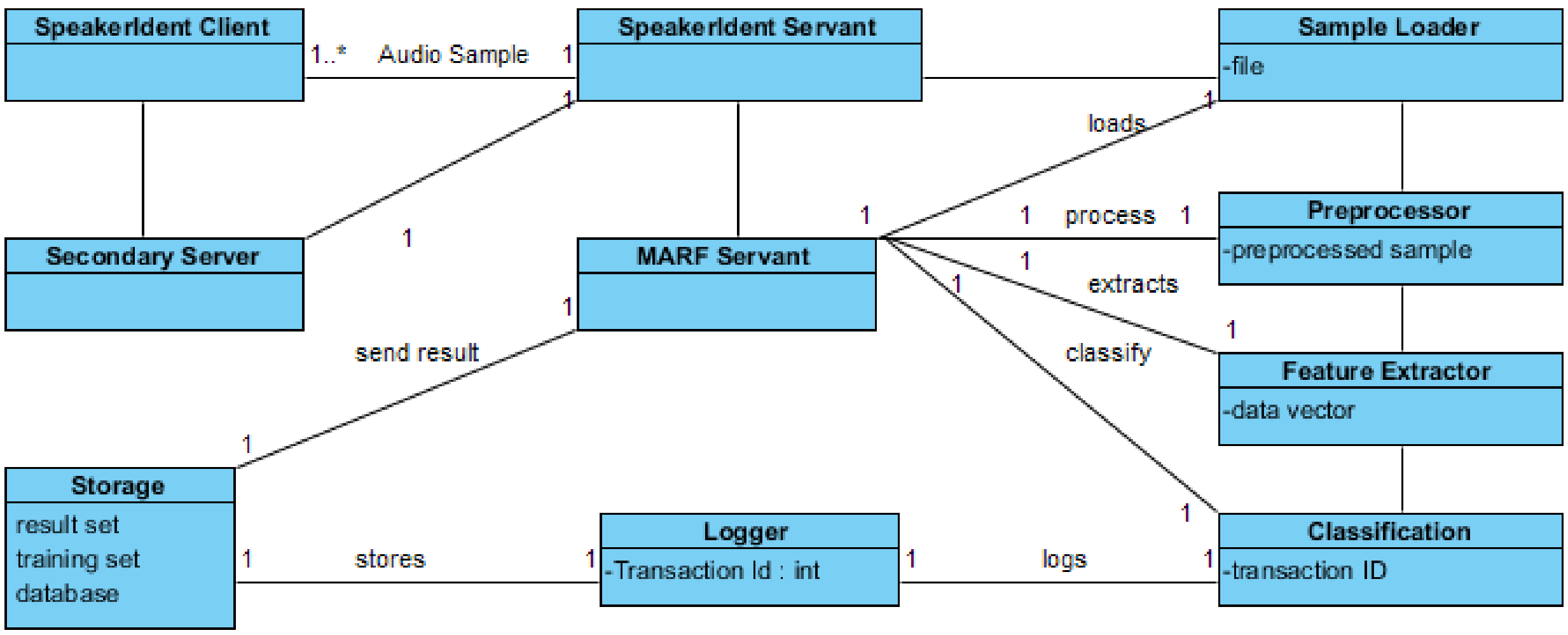

*Figure 11: Conceptual Model of DMARF*

b) ***GIPSY** Conceptual Model*

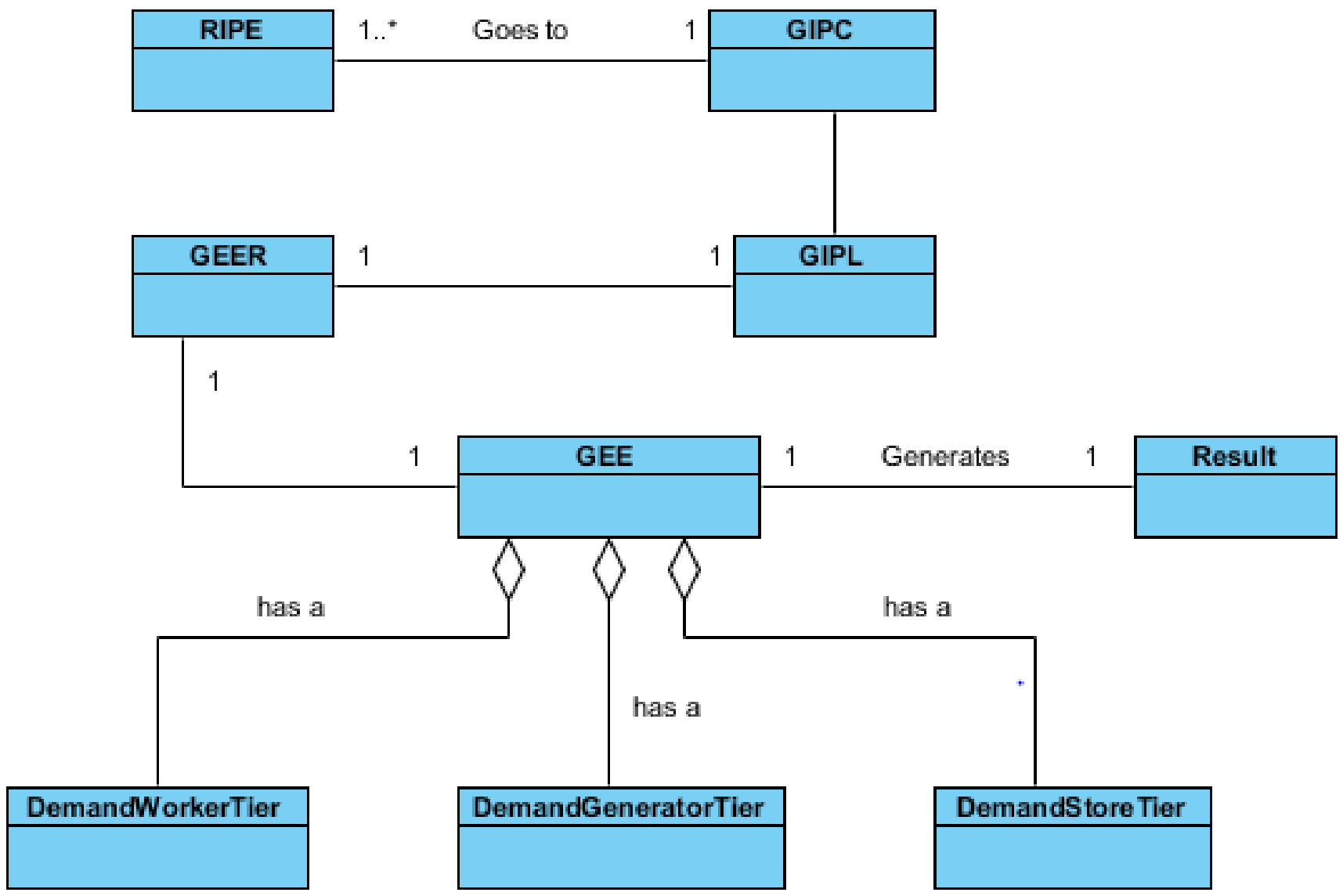

*Figure 12: Conceptual Model of GIPSY*

In the above fig. 12 the user gives input to the system. The GIPSY program first goes to the complier that is GIPC. Then the program is translated into source independent program, GIPL applying translation rules. Further, a source language independent GEER is generated on the basis of GIPL program. GEER then passes it to GEE which consists of different tiers (Demand Storage Tier, Demand Worker Tier, and Demand Generator Tier). The program is executed in GEE and the result is obtained.

c) *Fused DMARF-Over GIPSY Runtime Architecture**(DOGRTA)***

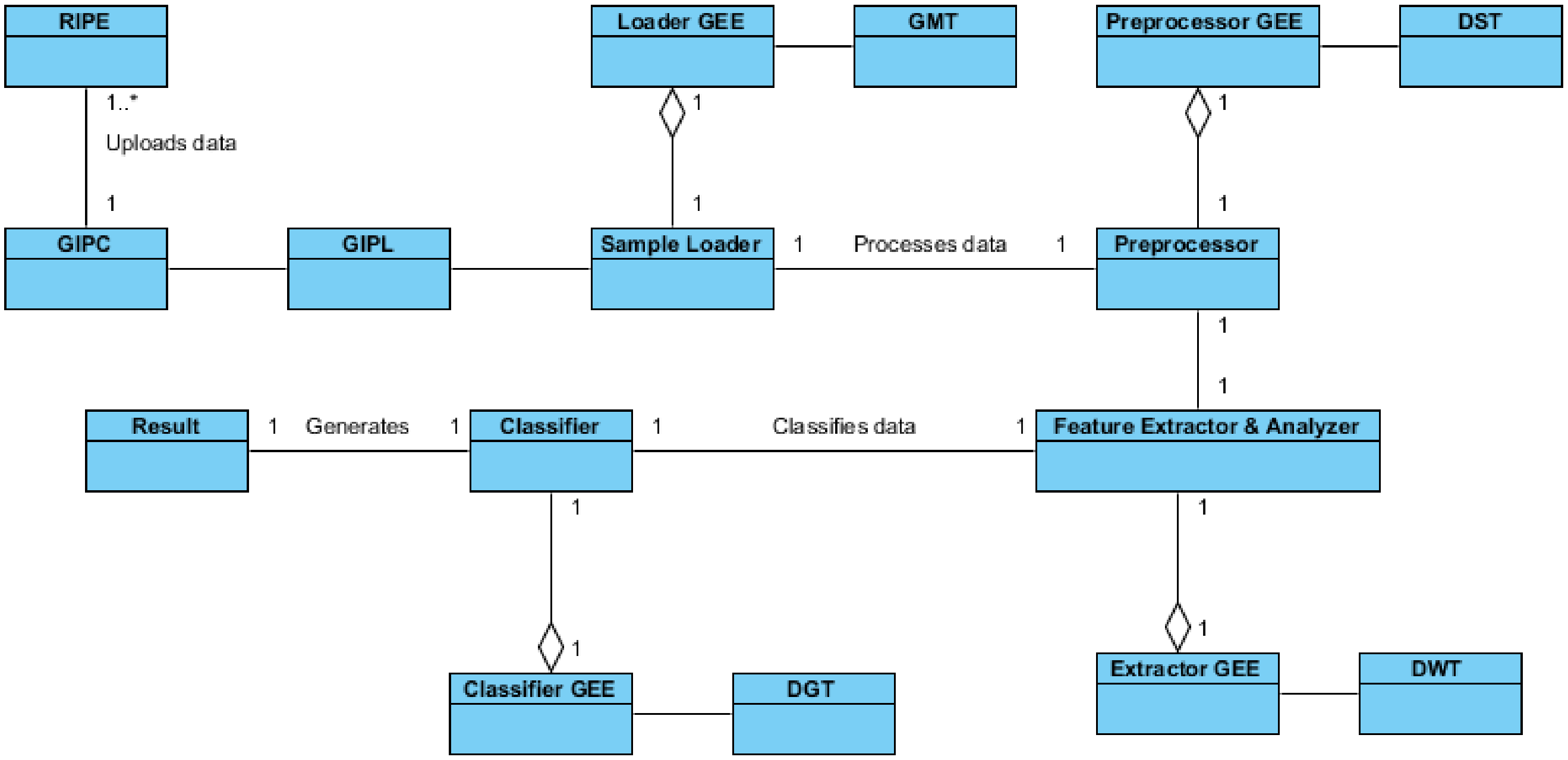

*Figure 13:* Fused DMARF-Over GIPSY Runtime Architecture **(DOGRTA)** Conceptual Model

The above merged domain model shows how the GEE multi-tier architecture of GIPSY is used by all the pipelined stages of DMARF. GIPSY follows a demand driven eductive execution model i.e. General Eduction Engine (GEE) which evaluates intentional expressions. In this model, the atomicity feature is provided to DMARF at runtime using General Eduction Engine (GEE) multi-tier architecture. The procedural demand is generated, delivered to a network demand store. The generated procedural demand is picked up by an observer located on other node for evaluation purpose. As soon as the evaluation completes the computation result is stored. It shows the implementation of distributed asynchronous communication.

*III. Actual Architecture UML Diagrams*

*1) Actual Architecture UML Diagrams of DMARF and GIPSY*

*a) **DMARF** UML Class Diagram*

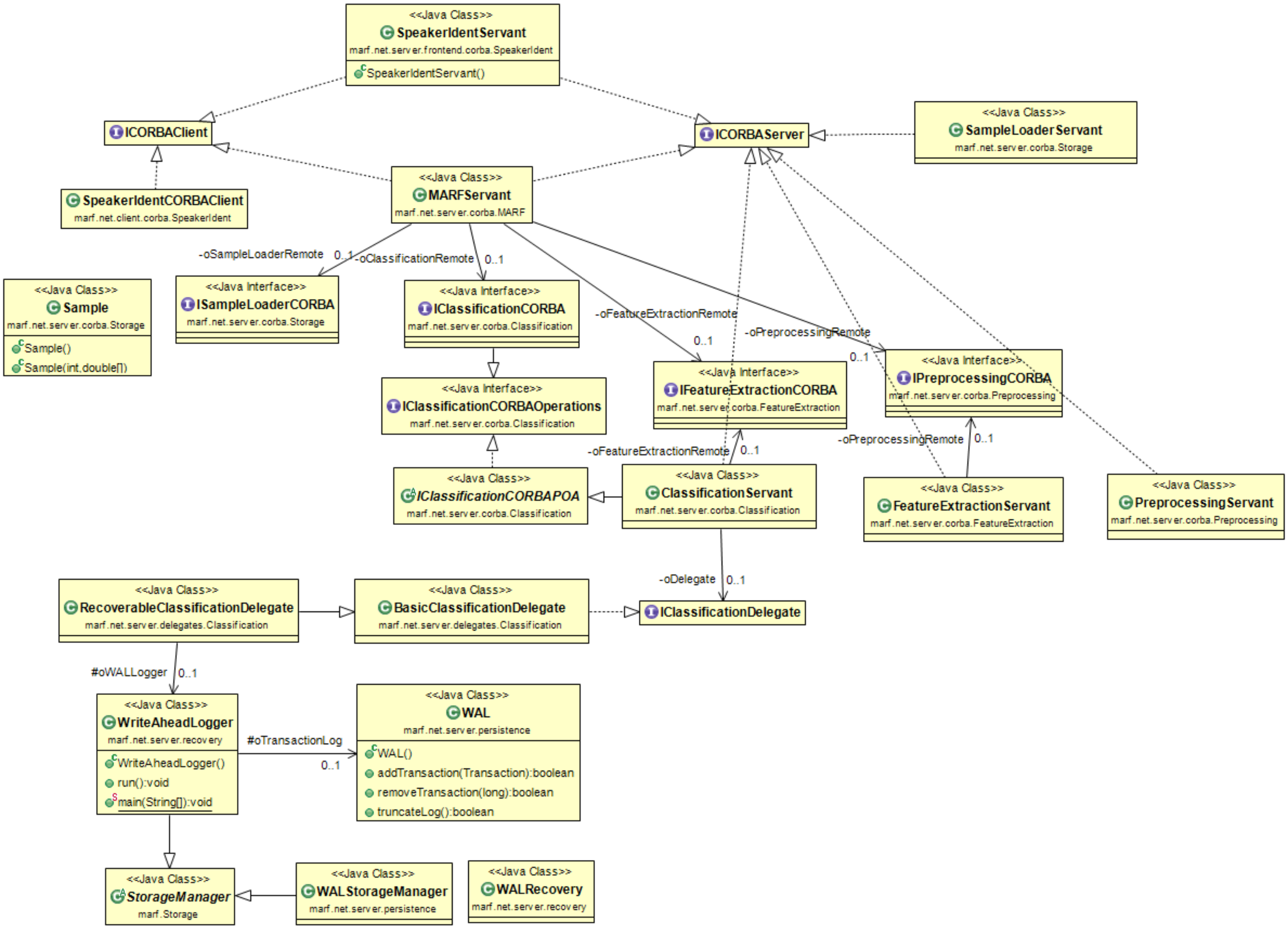

*Figure 14: UML Class Diagram of DMARF*

The entire design is summarized in two class diagrams shown in Fig 14 and Fig 14.1 representing the major modules and their relationships. The diagram shows only the CORBA details (and RMI and WS are similar, but omitted from this diagram). The class diagram shows stages of the pipeline as services as well as stages like sample loading, front-end application service (e.g. speaker identification service, etc.). It indicates different levels of basic front-ends, from higher to lower which client applications may invoke as well as services may invoke other services through their front-ends while executing in the pipeline model. At the beginning of the hierarchy are the ICORBAClient and ICORBAServer are independent of a communication technology type of interfaces that “mark” the would-be classes of either type and they all are connected to MARF servant. The pipelines sample loader, processor, feature extractor and classification are implementing their respective Interfaces like IPreprocessingCORBA, IFeatureExtractionCORBA, IClassificationCORBA which are communicating to MARF Servant. Output of Classification is result set and training set which is stored in file or in a database. Finally, on the server side, the RecoverableClassificationDelegate interacts with the WriteAheadLogger for transaction information. The storage manager here serializes the WAL entries. The Database contains stats of classification and is only written by the SpeakerIdent front-end. All, Database, Sample, Result, and ResultSet and TrainingSet needs to implement Serializable (java.io) to be able to be stored on disk or transferred over a network. The class diagram does not show serializable class as it's not actually present in DMARF code, but in .jar file.

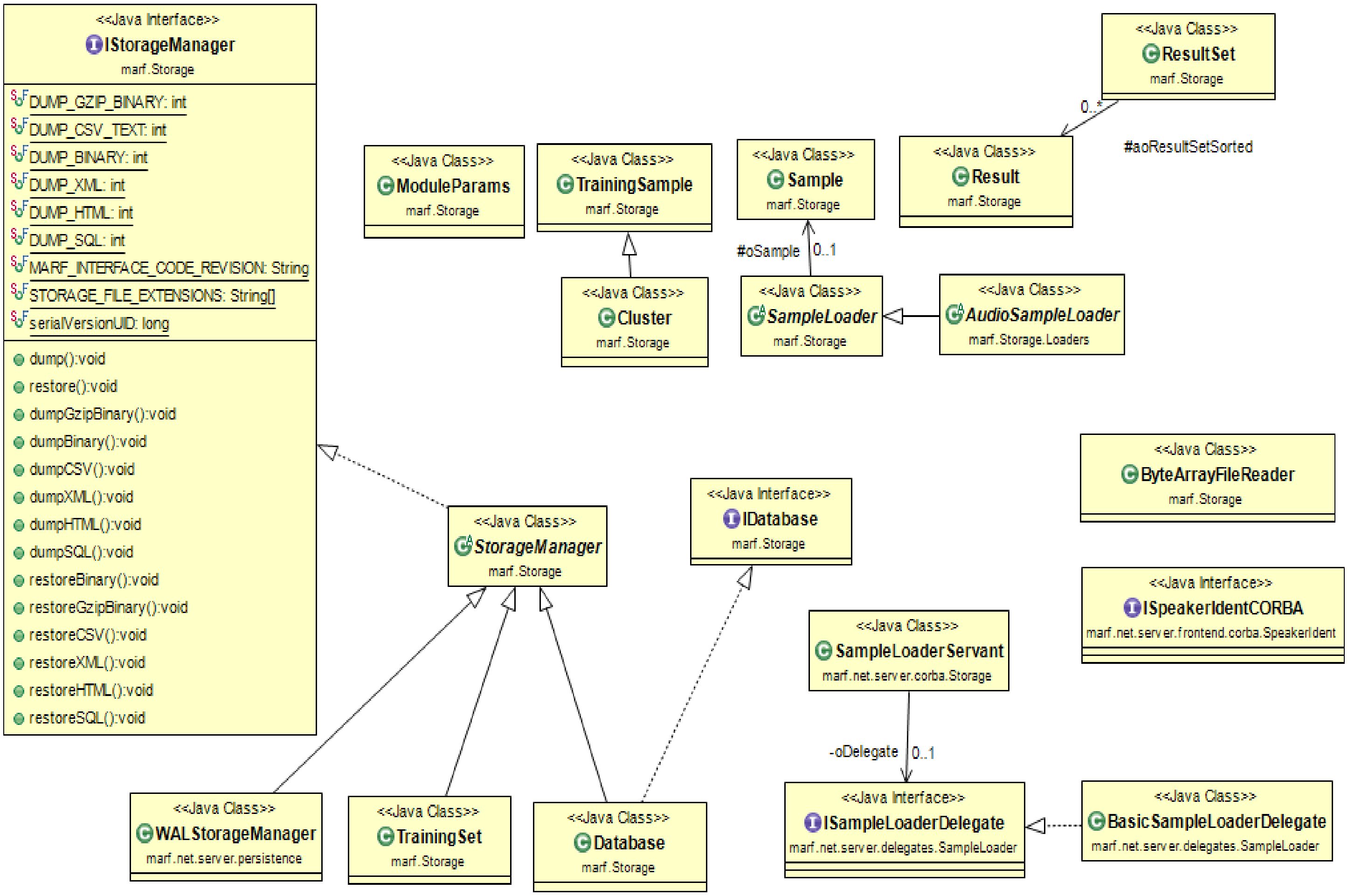

*Figure 14.1: UML Class Diagram of DMARF- Storage Module*

*b) **GIPSY** UML Class Diagram*

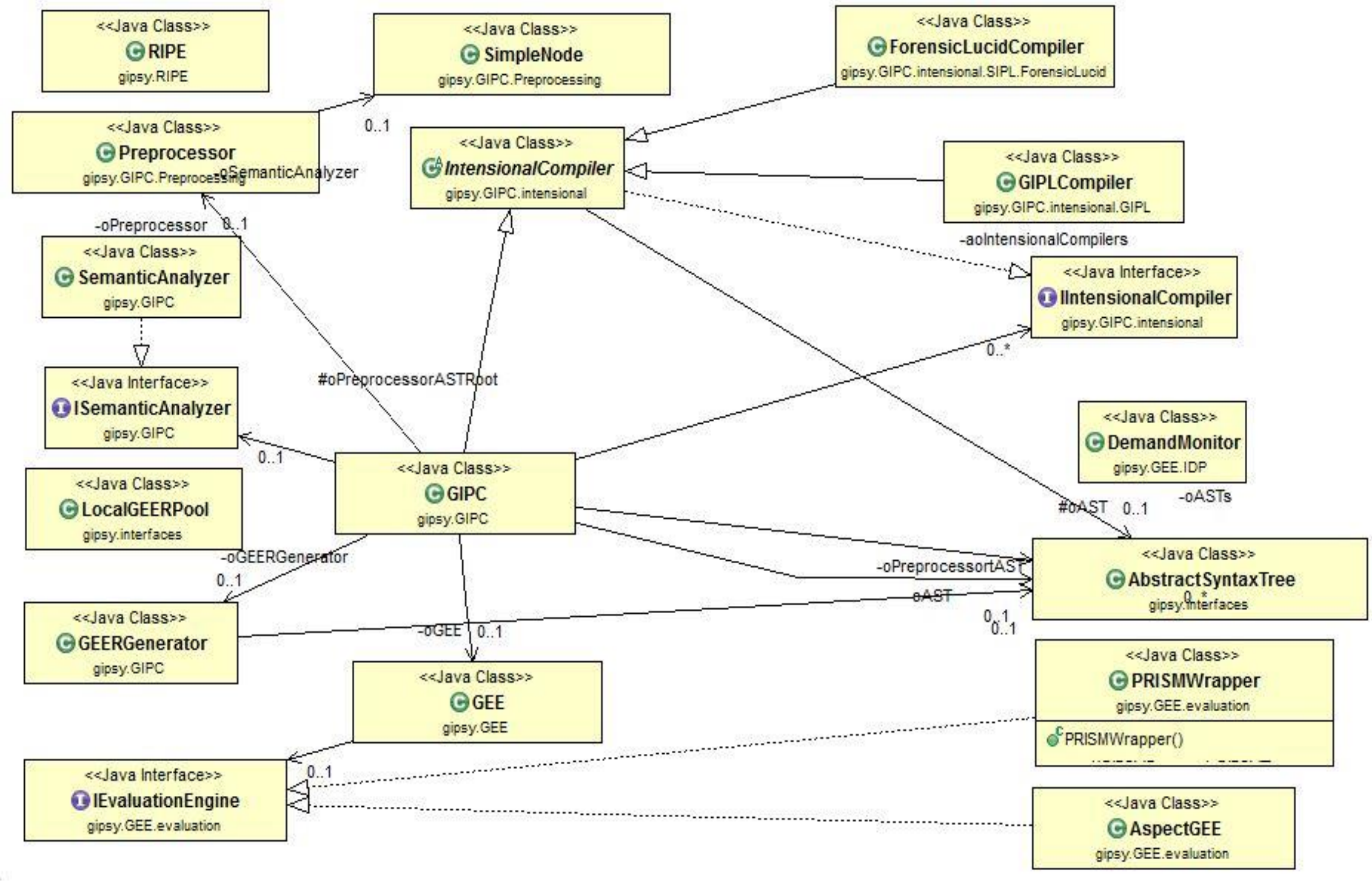

*Figure 15 UML Class Diagram of GIPSY*

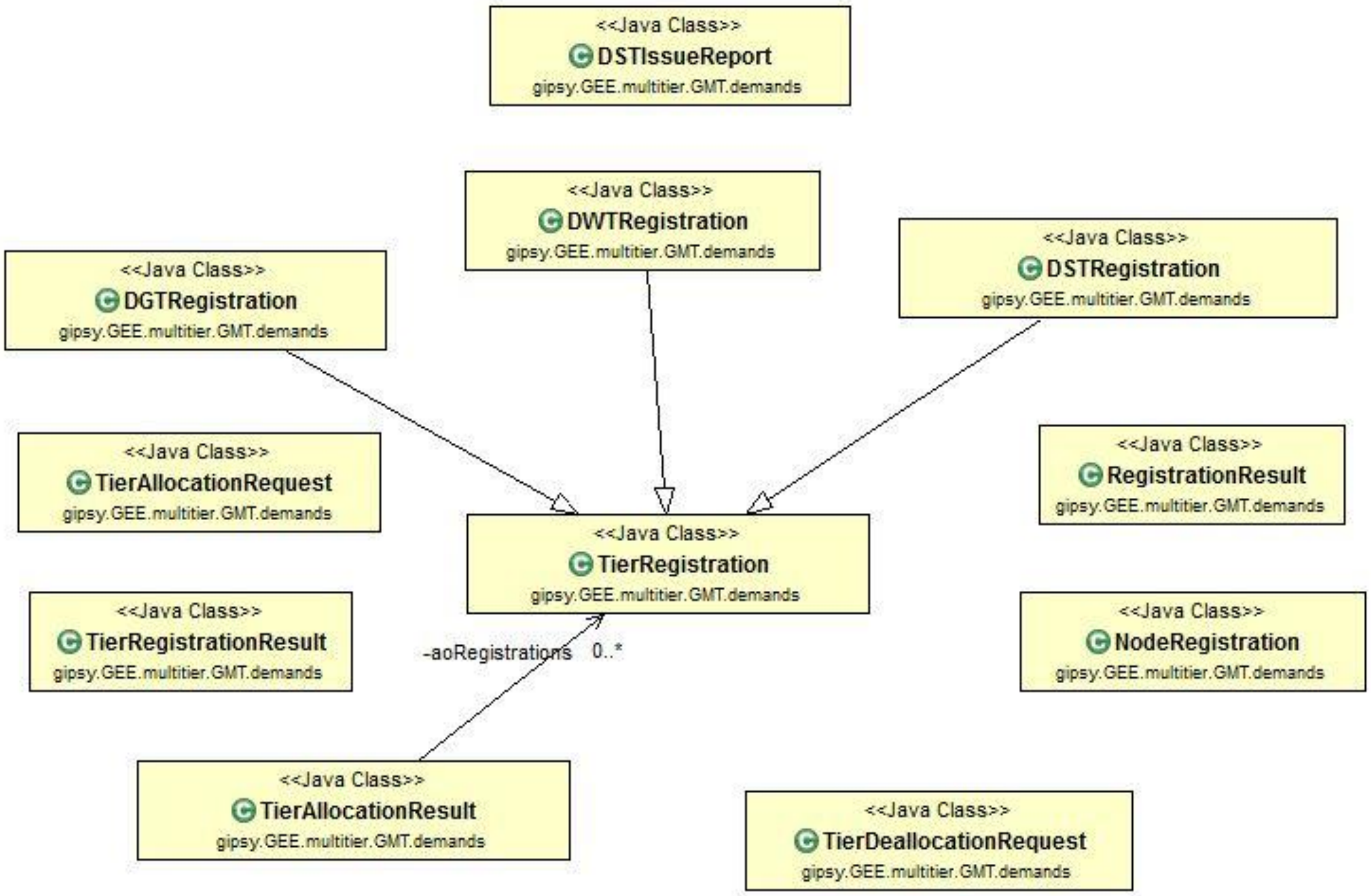

*Figure 15.1: UML Class Diagram of GIPSY- Multi-tier Architecture*

The Fig. 15 and 15.1 are the class diagram of the GIPSY, which represents the relationship between the classes of the GIPSY. The classes depicted in the class diagram are RIPE, SimpleNode, GIPC, Preprocessor, LocalGEERPool, SemanticAnalyzer, ISemanticAnalyzer, IntentionalComplier, IIntentionalComplier, ForensicLucidCompiler, GEERGenerator, AbstractSyntaxTree, GIPLCompiler, AspectGEE, GEE, DemandMonitor, IEvaluationEngine, PRISMWrapper, DSTIssueReport,DGTRegistration, DWTRegistration, DSTRegistration, TierAllocationRequest, TierRegistration,RegistrationResult, NodeRegistration, TierDeallocationRequest and TierRegistrationResult.

There are various relationships between the mentioned classes. The RIPE class is the environment for the user at run-time which is an interface to the user. SimpleNode is that class which can be a computer through which the user can access the GIPSY Features. The GIPC is associated with Preprocessor, GEE, ISemanticAnalyzer, AbstractSyntaxTree and GEERGenerator. Moreover, GIPC is inherited in IntensionalCompiler and GIPLComplier is inherited in IntensionalCompiler. Further, the GIPLComplier is associated with SemanticAnalyzer. The tier allocation process is done through GMTController, NodeRegistartion,TierAllocationRequest,DSTController, DGTController and the DWTController. The GMTController issues and allocates the new tier after getting the TierAllocationRequest. The Tier Deallocation Process is done through TierDeallocationRequest, TierDeallocationResult.

GEE evaluates the result using three classes i.e. AspectGEE, PRISMWrapperIEvaluationEngine. Moreover, the AspectGEE, PRISMWrapper is inherited in IEvaluationEngine.

### 2) *Diagrams and the relationships between the classes*

#### *a) DMARF*

a.1. The main differences between the conceptual architecture and the actual architecture:-

- In conceptual architecture, the client interacts with the DMARF system through the server and provides data sample to sample loader for further pattern processing whereas, on the other side, in actual architecture, the client interacts with the DMARF system through the SpeakerIdentServer in which client and server divided into three interfaces each and these are IRMIClient, CORBAclient, IWSclient, IRMIserver, CORBAserver, IWSserver.
- In conceptual architecture, logger maintains the message log and holds the transaction ids and store it in database for recovery whereas, on the other side, in actual architecture, RecoverableClassificationDeligate extracts the transaction information through WriteAheadLogger.
- In conceptual architecture, preprocessor and FeatureExtractorAnalyzer extracts the data through sampleLoader whereas, on the other side, in actual architecture, preprocessor and FeatureExtractorAnalyzer extract the data directly through the server.
- In conceptual architecture, DMARF servant gets the results from the database on the bases of instruction set and training set whereas, on the other side, in actual architecture, the results from database are stored in a disk on the bases of instruction set and training set that implements the serializable form and transfers over a network.
- In conceptual architecture, secondary server is provided incase if the primary server goes down due to some catastrophic condition, secondary server will takes up the charge and whole of the load will be handled by secondary server whereas in actual architecture, secondary server is not included.

a.2. Conceptual and Actual Class Mapping

- Client and Server class of conceptual domain model mapped with the actual IClient and IServer class diagram of DMARF, which enables to choose
- Communication technology type either manually thorough client or automatic through server.
- Sample loader both in conceptual and actual class diagram sample will get loaded into the sample loader for further pattern processing.
- Logger from conceptual model mapped with WriteAheadLogger of actual class diagram to maintain the message log.
- Database of conceptual class mapped with StorageManager of actual class diagram to store the log messages for recovery purposes.

a.3. Discrepancy between the concepts and the actual classes

The domain diagram has secondary server, but in the actual class diagram there is not concept of secondary server. Domain diagram is irrespective of language so there is no concept of Interfaces but as DMARF is a Java based project, so in class diagrams several Interfaces have been implemented in order to overcome the problem of no multiple Inheritance in Java. Some class diagrams have more attributes as compared to domain diagram. In the class diagram it is clear that every component has a specific function to perform. This also gives us the freedom to reuse the classes when necessary, reducing the coupling and enhancing the cohesion.

a.4. Reverse Engineering Tool used

For reconstruction of UML class diagram ObjectAid UML Explorer tool is used. The tool has been installed as a plugin in eclipse. The class diagram is created by dragging and dropping required classes into .ucls files. The tool makes it very easy to construct class diagram from a huge set of code. The ObjectAid UML Explorer tool also makes a copy of .ucml file in the form of .png file.

#### *b) GIPSY*

*b.1.* The main differences between the conceptual architecture and the actual architecture:-

In the conceptual architectures Model, we have not mentioned any node registration through GMT which handles the tier allocation and tier deallocation of the Demands. The domain model does not have Interfaces because it is conceptual but the class diagram has interfaces because its build by using classes in Java. The class diagram has attributes, declarations, methods which is not present in the domain conceptual classes.

*b.2.* Conceptual and Actual Class Mapping

In the Domain Model, the RIPE concept class corresponds to the RIPE class in the actual architecture. GIPC is equivalent to the GIPC class, GIPL parallels to the GIPLCompiler, GEER corresponds to the GEERGenerator ,GEE resembles to the GEE ,DemandWorkerTier corresponds to the DWTRegistration, DemandGeneratorTier relates to the DGTRegistration and DemandStoreTier matches to the DSTRegistration.

*b.3.* Discrepancy between the concepts and the actual classes

*b.4.* Reverse Engineering Tool used

ObjectAid UML Explorer tool is used for reconstruction of UML class diagram. The tool has been installed as a plug-in in eclipse. The class diagram is created by dragging and dropping required classes into .ucls files. The tool makes it very easy to construct class diagram from a huge set of code. The ObjectAid UML Explorer tool also makes a copy of .ucml file in the form of .png file, .jpeg file formats etc.

### *3) Rekatioship between Two Classes*

*a) DMARF*

Class WriteAheadLogger is using the WriteAheadLogger. An object of the class WriteAheadLogger has been called a used in recoverableClassficationDelegate

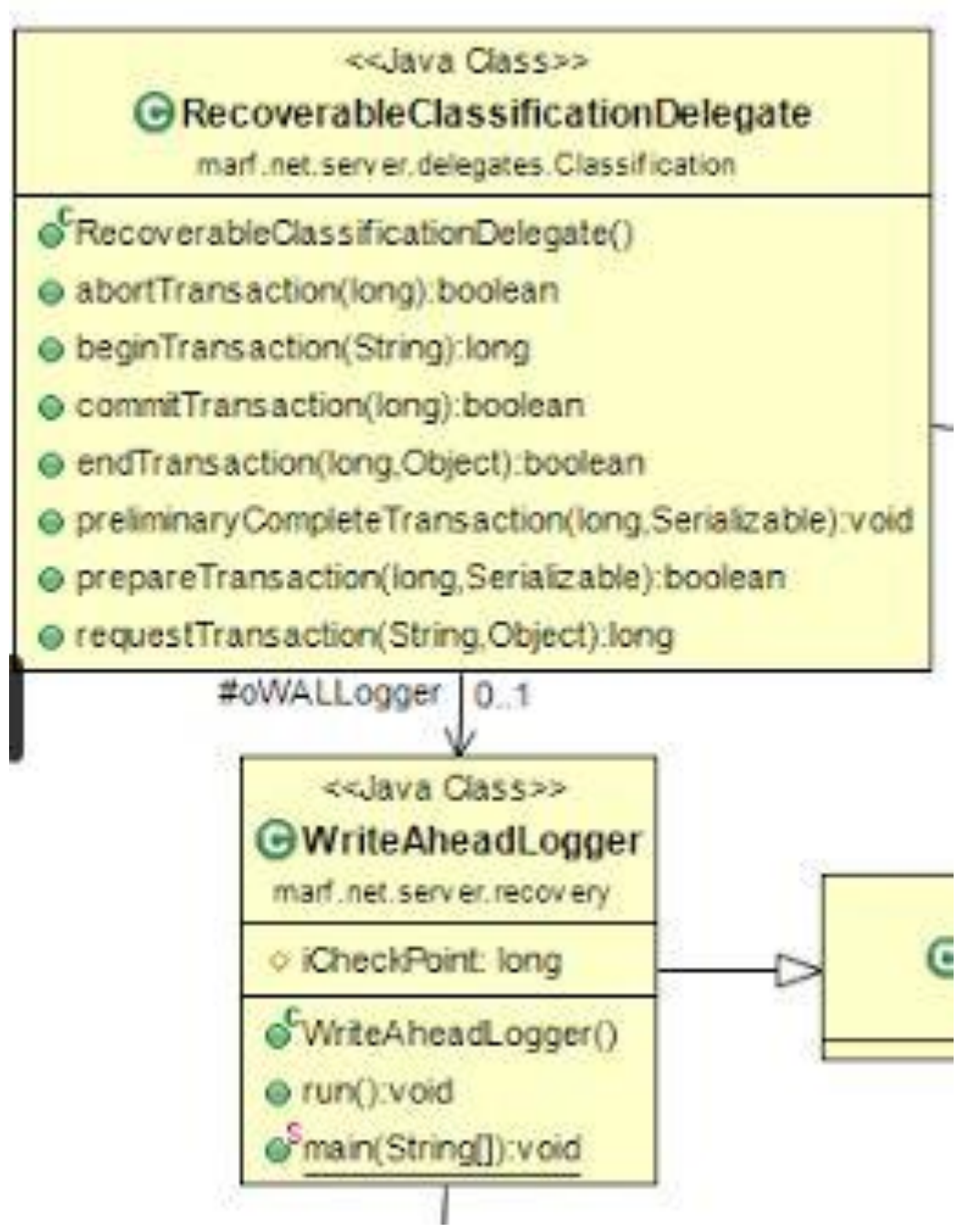

*Figure 16: Relationship between two classes of DMARF*

```
publicclassRecoverableClassificationDelegateextends
BasicClassificationDelegate
{
		protected WriteAheadLoggeroWALLogger = new
WriteAheadLogger();
		public RecoverableClassificationDelegate() throws
ClassificationException
		{
				super();
		}
		public booleanabortTransaction(long piTxnID)
		{
				returnsuper.abortTransaction(piTxnID);
		}
		public long beginTransaction(String pstrOperation) {
				return super.beginTransaction(pstrOperation);
		}
		public booleancommitTransaction(long piTxnID) {
				return super.commitTransaction(piTxnID);
		}
		public synchronized booleanendTransaction(long iTxnID, Object
poValue) {
				return super.endTransaction(iTxnID, poValue);
		}
		public void preliminaryCompleteTransaction(long plTxnID,
SerializablepoValue) {
				super.preliminaryCompleteTransaction(plTxnID,
poValue);
		}
		public booleanprepareTransaction(long piTxnID,
SerializablepoValue) {
				return super.prepareTransaction(piTxnID, poValue);
		}
		public synchronized long requestTransaction(String
pstrOperation, Object poValue) {
				return super.requestTransaction(pstrOperation,
poValue);
		}
}
```

```
public class WriteAheadLogger extends StorageManager implements
Runnable
{
		protected long iCheckPoint;
		protected WAL oTransactionLog = new WAL();
		public WriteAheadLogger() {
				super();
				this.oObjectToSerialize = this.oTransactionLog;
				this.iCurrentDumpMode = DUMP_GZIP_BINARY;
		}
		public void run()
		{
				throw new NotImplementedException();
		}
		public static void main(String[] args) {

		}
}
																}
}
```

*b) GIPSY*

The DGTRegistration and TierRegistration classes have the relationship in the GIPSY Framework. DGTRegistration is inherited in TierRegistration. Both classes are being used in the GEE for allocation of tiers to the users. The demands are generated, the classes in the GEE first registers the node in the node registration and then the process continues in the tiers, so both the tiers plays a part in the tiers of the GIPSY.

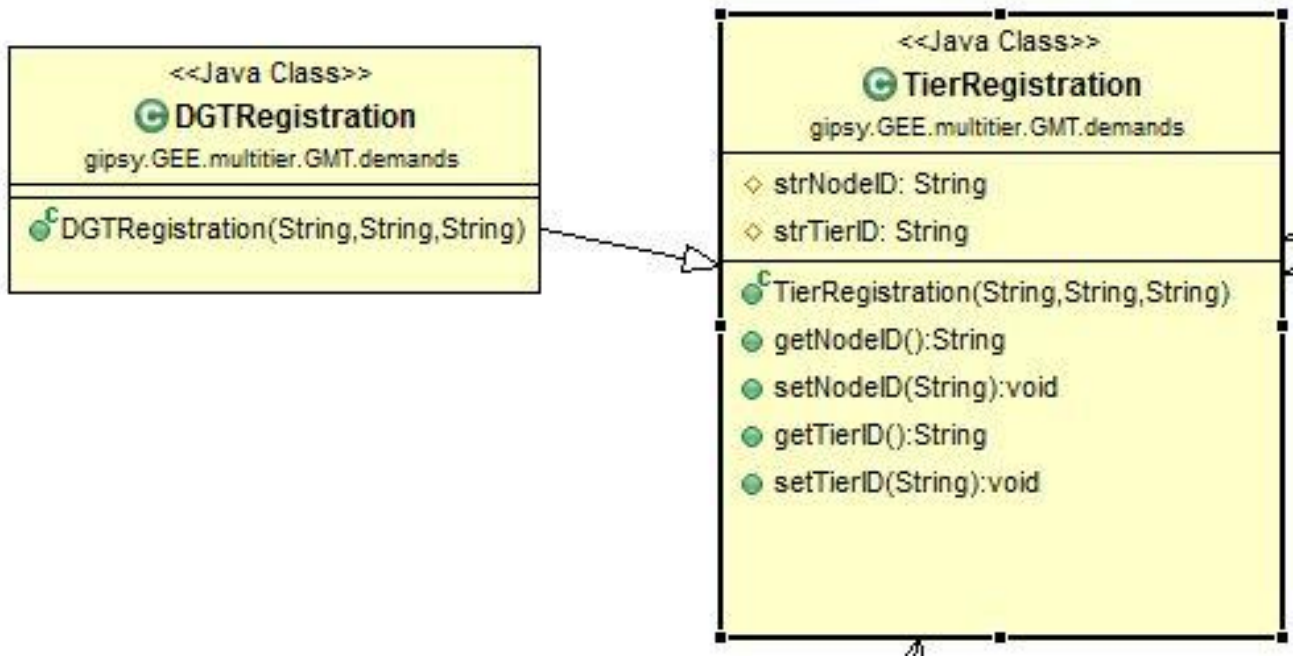

*Figure 17: Relationship between DGTRegistration and TierRegistration class of GIPSY*

```
package gipsy.GEE.multitier.GMT.demands;
import gipsy.GEE.IDP.demands.SystemDemand;
 * The registration demand contain information required for all tier re
public class TierRegistration
extends SystemDemand
{
    protected String strNodeID;
    protected String strTierID;

     * @param pstrNodeID
    public TierRegistration(String pstrNodeID, String pstrTierID,
            String pstrGMTTierID)
    {
        super();
        this.strNodeID = pstrNodeID;
        this.strTierID = pstrTierID;
        this.oDestinationTierID = pstrGMTTierID;
    }
    public String getNodeID()
    {
        return strNodeID;
    }
     public void setNodeID(String pstrNodeID)
    {
        this.strNodeID = pstrNodeID;
    }

    public String getTierID()
    {
        return strTierID;
    }
```

```
/**
package gipsy.GEE.multitier.GMT.demands;

import java.util.UUID;

 * The registration demand for either DGT or DWT registration
public class DGTRegistration
extends TierRegistration
{


     * @param pstrNodeID
    public DGTRegistration(String pstrNodeID, String pstrTierID,
            String pstrGMTTierID)
     {
        super(pstrNodeID, pstrTierID, pstrGMTTierID);
        this.oSignature = new DemandSignature(UUID.randomUUID().toString());
     }
}
```

## IV. Methodology

### 1) Refactoring

One of the inevitable negative effects of software evolution is design erosion. Refactoring is a technique that aims at counteracting this phenomenon by successively improving the design of software without changing its observable behaviour [23].

McCabe and Logiscope are two tools, used to identify and list code smells in the architecture on which refactoring can be done to improve the software architecture.

<u>*Logiscope*</u>

Maintainability is defined by Logiscope as "The capability of the software product to be modified. Modifications may include corrections, improvements or adaptation of the software to changes in environment, and in requirements and functional specifications [ISO/IEC 9126-1:2001]."

The Logiscope formula for maintainability defined as is:

MAINTAINABILITY = ANALYZABILITY+CHANGEABILITY+STABILITY+TESTABILITY

The Logiscope formulas for the class factor metrics are listed below:

1. ANALYZABILITY = cl_wmc + cl_comf + in_bases + cu_cdused
2. CHANGEABILITY = cl_stat + cl_func + cl_data
3. STABILITY = cl_data_publ + cu_cdusers + in_noc + cl_func_publ
4. TESTABILITY = cl_wmc + cl_func + cu_cdused

The Logiscope definitions of the operands in the MAINTAINABILITY formula are described below:

1. ANALYZABILITY measures the capability of the software product to be diagnosed for deficiencies or causes of failures in the software, or for the parts to be modified to be identified [ISO/IEC 9126-1:2001].
2. CHANGEABILITY measures the capability of the software product to enable a specified modification to be implemented [ISO/IEC 9126-1:2001].
3. STABILITY measures the capability of the software product to avoid unexpected effects from modifications of the software [ISO/IEC 9126-1:2001].
4. TESTABILITY measures the capability of the software product to enable modified software to be validated [ISO/IEC 9126-1:2001].

#### a) Extracted DMARF data

Collection of the maintainability measurement data on the DMARF code and results are shown in Fig 18.

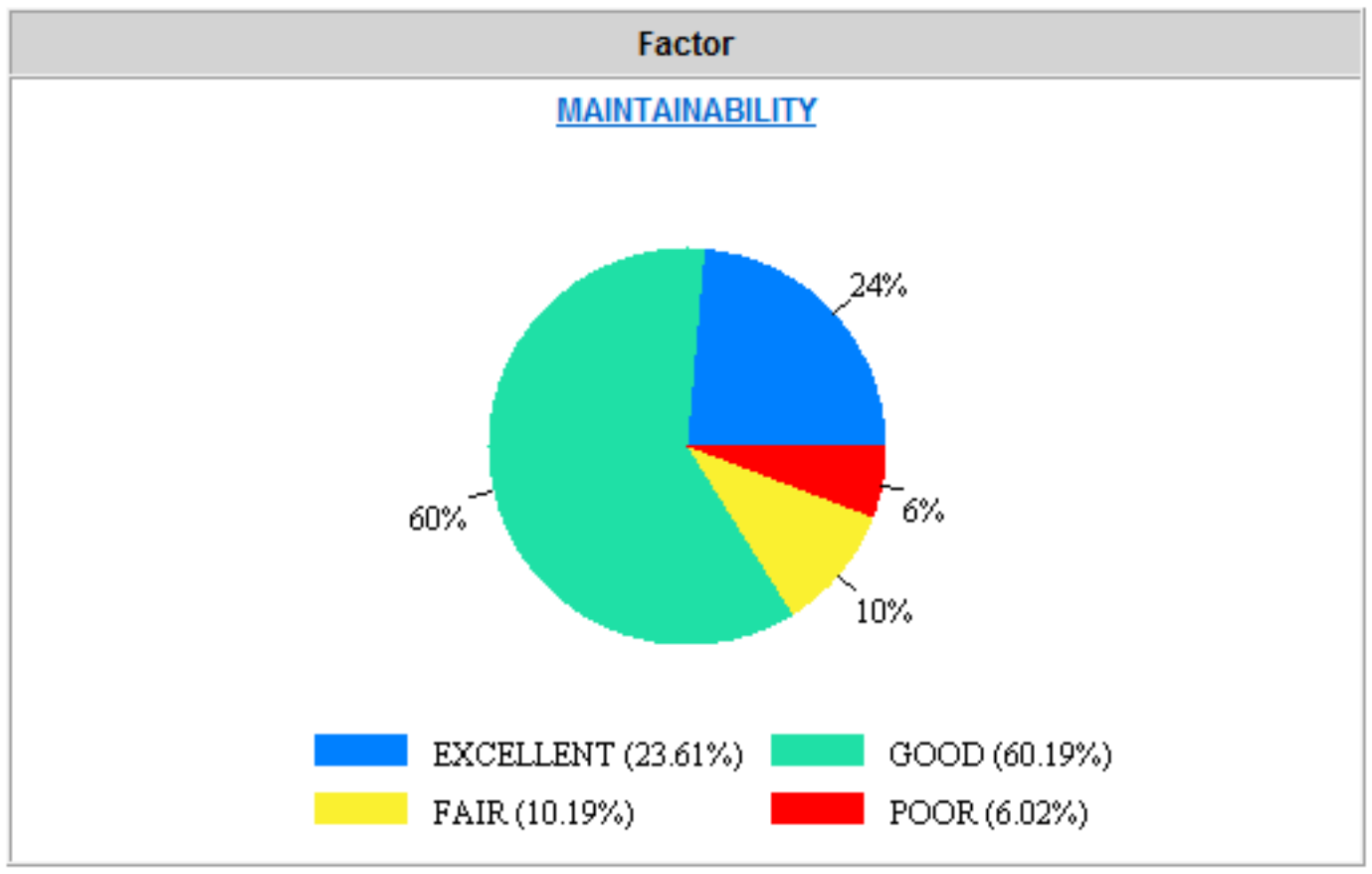

*Figure 18: DMARF MAINTAINABILITY at class factor level*

The maintainability of the MARF code as per class factor level is GOOD as observed from the above pie chart.

*b) Extracted GIPSY data*

Collection of the maintainability measurement data on the GIPSY Code and results are shown Fig.19 the Pie chart below:

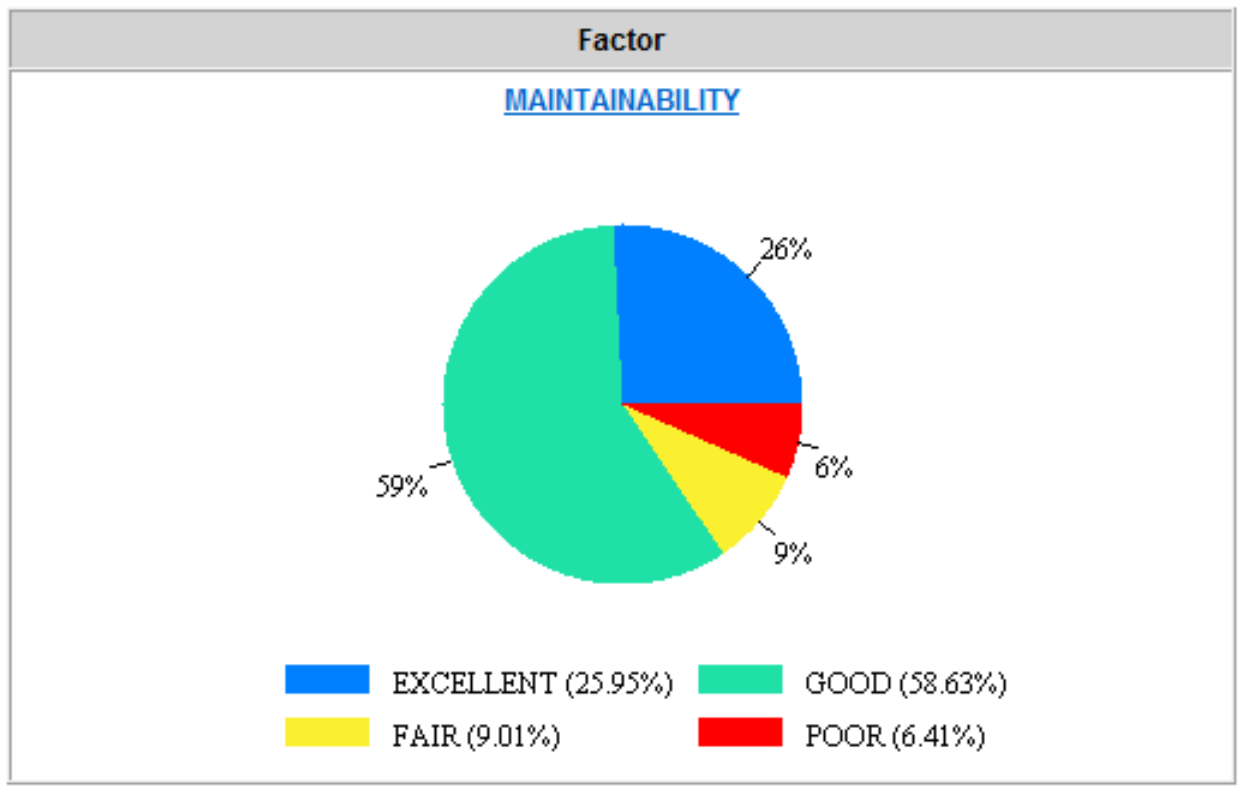

*Figure 19: GIPSY MAINTAINABILITY at class factor level*

The maintainability of the GIPSY code as per class factor level is GOOD.

<u>*McCabe*</u>
McCabe tools is used to find out the Cyclomatic complexity of the DMARF and GIPSY case studies in the form of metric. The Analysis of the quality of methods (modules) and classes of the DMARF and GIPSY case studies in terms of Average Cyclomatic Complexity [v(G)], Essential Complexity [ev(G)], Module Design Complexity [iv(G)] is mentioned in the subsequent sections:

*a) MARF*
The Analysis of the quality of methods (modules) of the DMARF case study in terms of Average Cyclomatic Complexity [v(G)], Essential Complexity [ev(G)], Module Design Complexity [iv(G)] in Fig. 20.

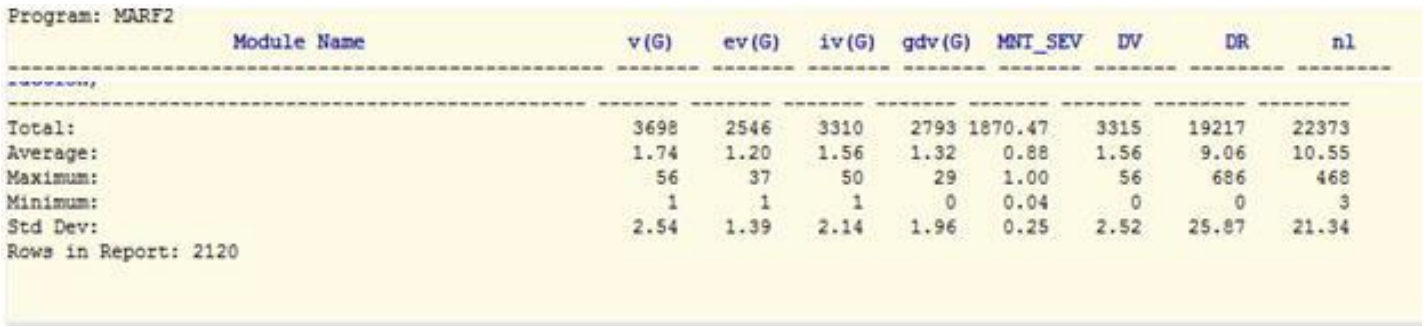

Program: MARF2

| Module Name | v(G) | ev(G) | iv(G) | gdv(G) | MNT_SEV | DV | DR | nl |
|---|---|---|---|---|---|---|---|---|
| Total: | 3698 | 2546 | 3310 | 2793 | 1870.47 | 3315 | 19217 | 22373 |
| Average: | 1.74 | 1.20 | 1.56 | 1.32 | 0.88 | 1.56 | 9.06 | 10.55 |
| Maximum: | 56 | 37 | 50 | 29 | 1.00 | 56 | 686 | 468 |
| Minimum: | 1 | 1 | 1 | 0 | 0.04 | 0 | 0 | 3 |
| Std Dev: | 2.54 | 1.39 | 2.14 | 1.96 | 0.25 | 2.52 | 25.87 | 21.34 |

Rows in Report: 2120

*Figure 20: DMARF complexity metrics from McCabe*

The Analysis of the quality of Classes of the DMARF and GIPSY case studies in terms of Average Coupling Between Objects [CBO], Weighted Methods per Class [WMC], Response For Class [RFC], Depth of Inheritance Tree [DIT/Depth], Number of Children [NOC] shown in Fig. 21

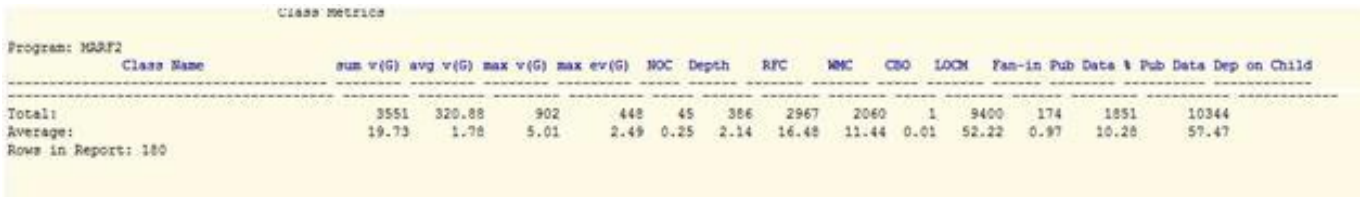

Class Metrics

Program: MARF2

| Class Name | sum v(G) | avg v(G) | max v(G) | max ev(G) | NOC | Depth | RFC | WMC | CBO | LOCM | Fan-in | Pub Data | % Pub Data | Dep on Child |
|---|---|---|---|---|---|---|---|---|---|---|---|---|---|---|
| Total: | 3551 | 320.88 | 902 | 448 | 45 | 386 | 2967 | 2060 | 1 | 9400 | 174 | 1851 | 10344 | |
| Average: | 19.73 | 1.78 | 5.01 | 2.49 | 0.25 | 2.14 | 16.48 | 11.44 | 0.01 | 52.22 | 0.97 | 10.28 | 57.47 | |

Rows in Report: 180

*Figure 21: DMARF class metrics from McCabe*

TABLE IX. ANALYSIS OF DMARF MCCABE METRICS

| Parameter/metric | Analyzed value | Threshold | Analysis |
|---|---|---|---|
| V(G) | 1.74 | 10.0 | Lesser analyzed value indicates that the design has less nested functions and methods. |
| ev(G) | 1.20 | 4.0 | The analyzed value of essential complexity is lesser than the threshold. This indicates that the given design is using structured programming constructs. |
| Iv(G) | 1.56 | 7.0 | The lesser value of module design complexity indicates that lesser decisions are involved in subroutine calls making this design a reliable and its integration with other modules and integration testability is easier. |
| CBO | 1.0 | 2.0 | Analyzed value is less than the threshold, it indicates that the design is easy to maintain, reuse, and test. Lesser value also indicates that the classes are loosely coupled. |
| WMC | 2060 | 14.0 | Higher value than threshold indicates high design complexity and hard to maintain and reuse. |
| RFC | 2967 | 100.0 | High value of RFC than the threshold indicates high design complexity and difficulty in debugging. |
| NOC | 386 | 7.0 | High value of analyzed value of NOC indicates high reuse of base class and hence higher could be the probability of propagating the bug in subclasses. |
| DIT | 45 | 3.0 | Since the analyzed value is much higher than the threshold, it indicates that the design is highly complex and high maintenance effort is required. |
| | | | could be the probability of propagating the bug in subclasses. |

*b) GIPSY*

The Analysis of the quality of methods (modules) of the GIPSY case study in terms of Average Cyclomatic Complexity [v(G)], Essential Complexity [ev(G)], Module Design Complexity [iv(G)]in Fig.22.

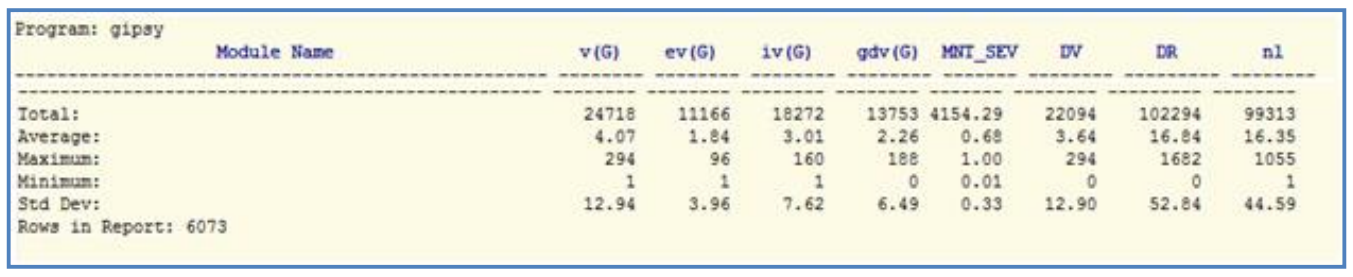

Program: gipsy

| Module Name | v(G) | ev(G) | iv(G) | gdv(G) | MNT_SEV | DV | DR | nl |
|---|---|---|---|---|---|---|---|---|
| Total: | 24718 | 11166 | 18272 | 13753 | 4154.29 | 22094 | 102294 | 99313 |
| Average: | 4.07 | 1.84 | 3.01 | 2.26 | 0.68 | 3.64 | 16.84 | 16.35 |
| Maximum: | 294 | 96 | 160 | 188 | 1.00 | 294 | 1682 | 1055 |
| Minimum: | 1 | 1 | 1 | 0 | 0.01 | 0 | 0 | 1 |
| Std Dev: | 12.94 | 3.96 | 7.62 | 6.49 | 0.33 | 12.90 | 52.84 | 44.59 |

Rows in Report: 6073

*Figure 22: GIPSY complexity metrics from McCabe*

The Analysis of the quality of Classes of the GIPSY case study in terms of Average Coupling Between Objects [CBO], Weighted Methods per Class [WMC], Response For Class [RFC], Depth of Inheritance Tree [DIT/Depth], Number of Children [NOC]shown in Fig. 23.

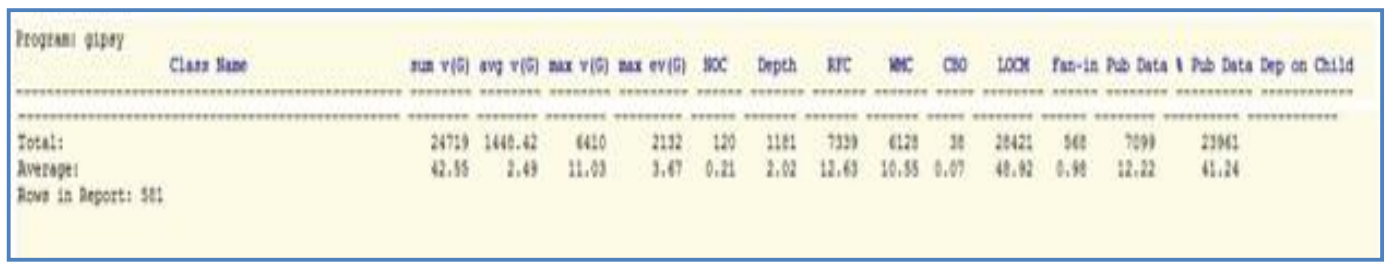

Program: gipsy

| Class Name | sum v(G) | avg v(G) | max v(G) | max ev(G) | NOC | Depth | RFC | WMC | CBO | LOCM | Fan-in | Pub Data | % Pub Data | Dep on Child |
|---|---|---|---|---|---|---|---|---|---|---|---|---|---|---|
| Total: | 24719 | 1440.42 | 6410 | 2132 | 120 | 1181 | 7339 | 6128 | 38 | 28421 | 568 | 7099 | 23961 | |
| Average: | 42.55 | 2.49 | 11.03 | 3.67 | 0.21 | 2.02 | 12.63 | 10.55 | 0.07 | 48.92 | 0.98 | 12.22 | 41.24 | |

Rows in Report: 581

*Figure 23: GIPSY class metrics from McCabe*

TABLE X. ANALYSIS OF GIPSY MCCABE METRICS

| **Parameter/metric** | **Analyzed value** | **Threshold** | **Analysis** |
|---|---|---|---|
| V(G) | 4.07 | 10.00 | Lesser analyzed value indicates that the design has less nested functions and methods. |
| ev(G) | 1.84 | 4.00 | The analyzed value of essential complexity is lesser than the threshold. This indicates that the given design is using structured programming constructs. |
| Iv(G) | 3.01 | 7.00 | The lesser value of module design complexity indicates that lesser decisions are involved in subroutine calls making this design a reliable and its integration with other modules and integration testability is easier. |
| CBO | 0.07 | 2.00 | Analyzed value is less than the threshold, it indicates that the design is easy to maintain, reuse, and test. Lesser value also indicates that the classes are loosely coupled. |
| WMC | 10.55 | 14.00 | Lesser value than threshold indicates less design complexity and easy to maintain and reuse. |
| RFC | 12.63 | 100.00 | Less value of RFC than the threshold indicates less design complexity and easiness in debugging. |
| DIT | 2.02 | 7.00 | Since the analyzed value is lesser than the threshold, it indicates that the design is less complex and lesser maintenance effort is required. |
| NOC | 0.21 | 3.00 | Less value of analyzed value of NOC indicates less reuse of base class and hence lesser |

Measurement Data Analysis

In order to rank the code in each case study according to their class factor and class criteria levels, we extracted the data from Logiscope, assigned each level a weight (EXCELLENT=4, GOOD=3, FAIR=2, POOR=1), multiplied the number of occurrences of each level category by its respective weight, and summed the results. For example, a class with three occurrences of EXCELLENT and two of FAIR would score (3*4 + 2*2) = 16.

By ordering the classes according to their scores, we could visually identify the relative overall quality rank of every class. Using this method, we were able to identify particular packages that tend to have lower quality code, and we identified the following packages as have the worst quality code in their respective case studies:

- MARF: *marf.nlp.Parsing*
- GIPSY: *GIPSY.intensional.SIPL*

TABLE XI: lists the classes in *marf.nlp.Parsing*that are characterized as either fair or poor at the factor or criteria level.

TABLE XI. POOR AND FAIR CLASSES IN *MARF.NLP.PARSING*

| **Class Name** |
|---|
| marf.nlp.Parsing.GenericLexicalAnalyzer |
| marf.nlp.Parsing.GrammarCompiler.Grammar |
| marf.nlp.Parsing.GrammarCompiler.GrammarAnalyzer |
| marf.nlp.Parsing.GrammarCompiler.GrammarCompiler |
| marf.nlp.Parsing.GrammarCompiler.GrammarTokenType |
| marf.nlp.Parsing.GrammarCompiler.ProbabilisticGrammarCompiler |
| marf.nlp.Parsing.LexicalAnalyzer |
| marf.nlp.Parsing.LexicalError |
| marf.nlp.Parsing.Parser |
| marf.nlp.Parsing.ProbabilisticParser |
| marf.nlp.Parsing.SymbolTable |
| marf.nlp.Parsing.SyntaxError |
| marf.nlp.Parsing.TokenSubType |
| marf.nlp.Parsing.TokenType |
| marf.nlp.Parsing.TransitionTable |

TABLE XII. POOR AND FAIR CLASSES IN *GIPSY.INTENSIONAL.SIPL*

| **Class Name** |
|---|
| gipsy.GIPC.intensional.SIPL.JOOIP.JavaCharStream |
| gipsy.GIPC.intensional.SIPL.ForensicLucid.ForensicLucidParser |
| gipsy.GIPC.intensional.SIPL.ForensicLucid.ForensicLucidParserTokenManager |
| gipsy.GIPC.intensional.SIPL.ForensicLucid.ForensicLucidSemanticAnalyzer |
| gipsy.GIPC.intensional.SIPL.IndexicalLucid.IndexicalLucidParser |
| gipsy.GIPC.intensional.SIPL.IndexicalLucid.IndexicalLucidParserTokenManager |
| gipsy.GIPC.intensional.SIPL.JLucid.JGIPLParser |
| gipsy.GIPC.intensional.SIPL.JLucid.JGIPLParserTokenManager |
| gipsy.GIPC.intensional.SIPL.JLucid.JIndexicalLucidParser |
| gipsy.GIPC.intensional.SIPL.JLucid.JIndexicalLucidParserTokenManager |
| gipsy.GIPC.intensional.SIPL.JOOIP.ast.visitor.GenericVisitor |
| gipsy.GIPC.intensional.SIPL.JOOIP.ast.visitor.VoidVisitor |
| gipsy.GIPC.intensional.SIPL.JOOIP.JavaParser |
| gipsy.GIPC.intensional.SIPL.Lucx.LucxParser |
| gipsy.GIPC.intensional.SIPL.Lucx.LucxParserTokenManager |
| gipsy.GIPC.intensional.SIPL.ObjectiveLucid.ObjectiveGIPLParser |
| gipsy.GIPC.intensional.SIPL.ObjectiveLucid.ObjectiveGIPLParserTokenManager |
| gipsy.GIPC.intensional.SIPL.ObjectiveLucid.ObjectiveIndexicalLucidParser |
| gipsy.GIPC.intensional.SIPL.ObjectiveLucid.ObjectiveIndexicalLucidParserTokenManager |
| gipsy.GIPC.intensional.SIPL.JOOIP.ast.visitor.DumpVisitor |
| gipsy.GIPC.intensional.SIPL.IndexicalLucid.IndexicalLucidParserTreeConstants |
| gipsy.GIPC.intensional.SIPL.JOOIP.JavaParserTokenManager |
| gipsy.GIPC.intensional.SIPL.JOOIP.JOOIPCompiler |
| gipsy.GIPC.intensional.SIPL.JOOIP.ast.body.TypeDeclaration |
| gipsy.GIPC.intensional.SIPL.JOOIP.ast.expr.StringLiteralExpr |
| gipsy.GIPC.intensional.SIPL.JOOIP.ast.Node |
| gipsy.GIPC.intensional.SIPL.JLucid.JLucidCompiler |
| gipsy.GIPC.intensional.SIPL.JOOIP.ast.body.BodyDeclaration |
| gipsy.GIPC.intensional.SIPL.JOOIP.ast.body.VariableDeclaratorId |
| gipsy.GIPC.intensional.SIPL.JOOIP.ast.expr.CharLiteralExpr |
| gipsy.GIPC.intensional.SIPL.JOOIP.ast.expr.DoubleLiteralExpr |
| gipsy.GIPC.intensional.SIPL.JOOIP.ast.expr.Expression |
| gipsy.GIPC.intensional.SIPL.JOOIP.ast.expr.IntegerLiteralMinValueExpr |
| gipsy.GIPC.intensional.SIPL.JOOIP.ast.expr.LongLiteralMinValueExpr |
| gipsy.GIPC.intensional.SIPL.JOOIP.ast.expr.NameExpr |
| gipsy.GIPC.intensional.SIPL.JOOIP.ast.stmt.BlockStmt |
| gipsy.GIPC.intensional.SIPL.JOOIP.ast.stmt.Statement |
| gipsy.GIPC.intensional.SIPL.JOOIP.ast.type.Type |
| gipsy.GIPC.intensional.SIPL.JOOIP.Token |

### *2) Identification of Code Smells and System Level Refactoring’s*

#### *a) DMARF*

MARF.java

Going through code in the bad classes we find classes can be re-factored based on the large number of attributes and methods in attributes located in single class.
This could be problematic class as it violates, single responsibility principle and it control large number of object implementing different functionality the solution.
We can to extract all the methods and fields, which are related to specific functionality into a separate class.

#### *b) GIPSY*

ObjectiveGIPLParser.java

For this particular class the number of public attributes are more. Therefore reducing the number of public attributes can result in higher security level of the class. It also increases the encapsulation. It can also be noted that the inheritance level for this is class is less as there are no children classes. The public methods can be made private or protected so that the security issues of the class are addressed.

### *3) Specific Refactoring’s to be Implemented in PM4.*

#### *a) DMARF*

**List of Source Code Smells**

Below is the list of smells and their definitions the code.

1) **GOD CLASS**: A god class violates, single responsibility principle and it control large number of object implementing different functionality the solution is to extract all the methods and fields, which are related to specific functionality into separate class.

   **Class**: NeuralNetwork.java
   (WMC=127, ATFD=83, TCC=0.0)

2) **Duplicate Code**: It is a sequence of source code that occurs more than once, either within a program or across different programs owned or maintained by the same entity. Duplicate code is generally considered undesirable for a number of reasons.

   **Class**: Configuration_SOAPSerializer

3) **Long Method:** A method function or procedure that has grown too large. A short term method is easier to read, easier to understand, and easier to troubleshoot.

   **Class**: MARFServant

Method: synchronized (sstrFileName)

4) **Dead Code**: Dead code is a code in the source code of a program which is executed but whose result is never used in any other computation. The execution of dead code wastes computation time as its results are never used.

**Class**: ClassificationException_SOAPSerializer

### V. *Identification of Design Patterns*

#### a) *DMARF*

#### 1. Singleton Pattern:

Singleton pattern is a design pattern that restricts the instantiation of a class to one object. This is useful when exactly one object is needed to coordinate actions across the system.

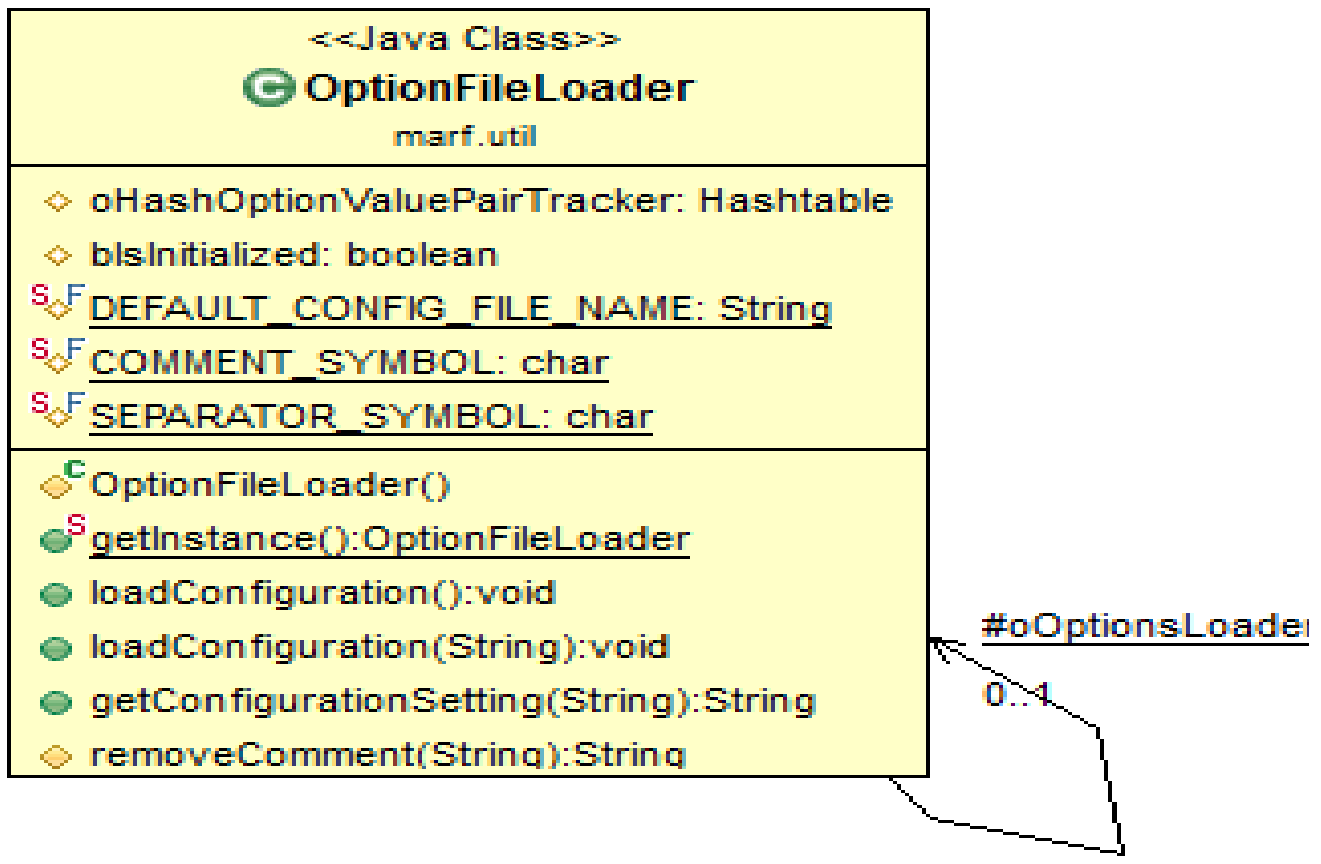

*Figure 24: Singleton Pattern*

```
public class OptionFileLoader{

protected static OptionFileLoader oOptionsLoaderInstance = null;


public static synchronized OptionFileLoader getInstance()
        {
                if(oOptionsLoaderInstance == null)
                {
                        oOptionsLoaderInstance = new OptionFileLoader();
                }

                Return oOptionsLoaderInstance;
        }
}
```

TABLE XIII. CODE SHOWING SINGLETON

```
public interface IClassification CORBA Operations
{
Boolean init
(marf.net.server.corba.Storage.ConfigurationpoConfig)
throws
marf.net.server.corba.CORBACommunicationException;

boolean classify() throws
marf.net.server.corba.CORBACommunicationException;
}
```

TABLE XIV. CODE SHOWING COMPONENT

#### 2. Decorator Pattern:

Decorator design allows behavior to be added to an individual object, either statically or dynamically, without affecting the behavior of other objects from the same class.

Component:
*IClassificationCORBAOperations*

Decorator
*IClassificationCORBAPOATie*

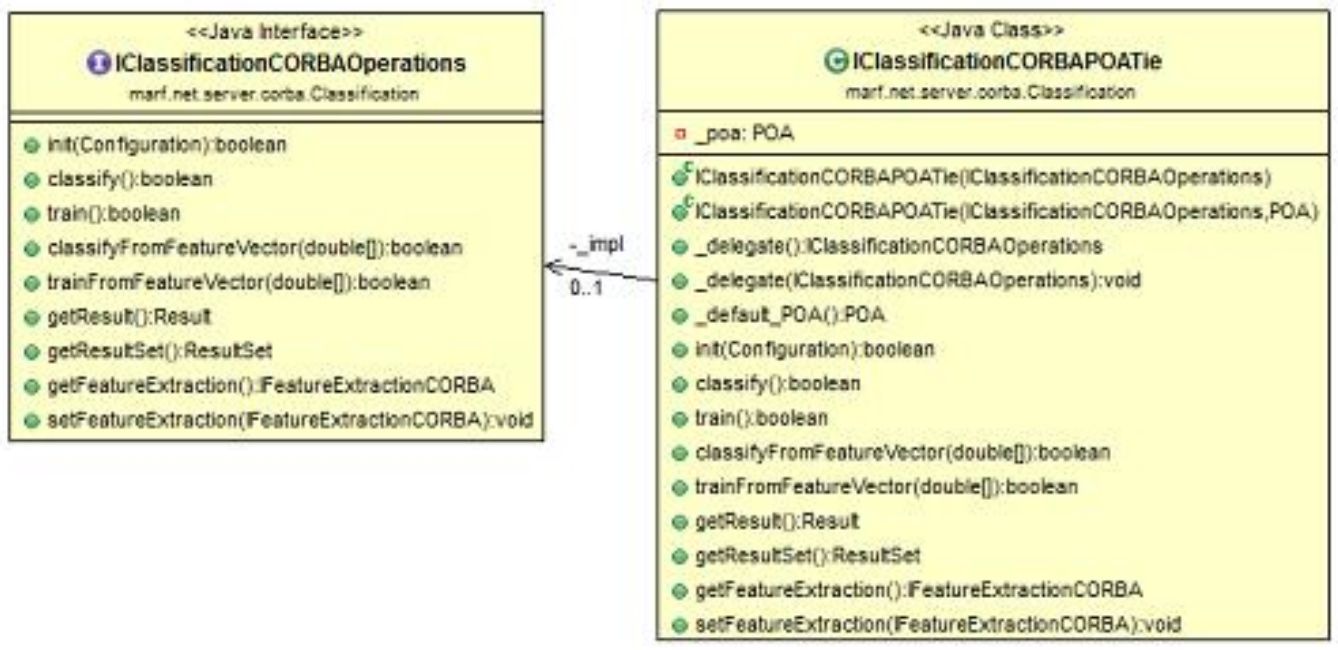

*Figure 25: Decorator Pattern*

```
publicclassIClassificationCORBAPOATieextendsIClass
ificationCORBAPOA
{……….

publicIClassificationCORBAPOATie(marf.net.server.cor
ba.Classification.IClassificationCORBAOperations
delegate ) {
this._impl = delegate;
publicbooleaninit(marf.net.server.corba.Storage.Configu
rationpoConfig)
throwsmarf.net.server.corba.CORBACommunicationExc
eption
{
return_impl.init(poConfig);
```

```
  }
publicboolean classify ()
throwsmarf.net.server.corba.CORBACommunicationException
  {
return_impl.classify();
  }

………}
```

TABLE XV. CODE SHOWING DECORATOR

### 3. Composite Pattern:

The composite pattern is a partitioning design pattern. The composite pattern describes that a group of objects are to be treated in the same way as a single instance of an object. The intent of a composite is to "compose" objects into tree structures to represent part-whole hierarchies.
Composite design pattern allows you to have a tree structure and ask each node in the tree structure to perform a task.

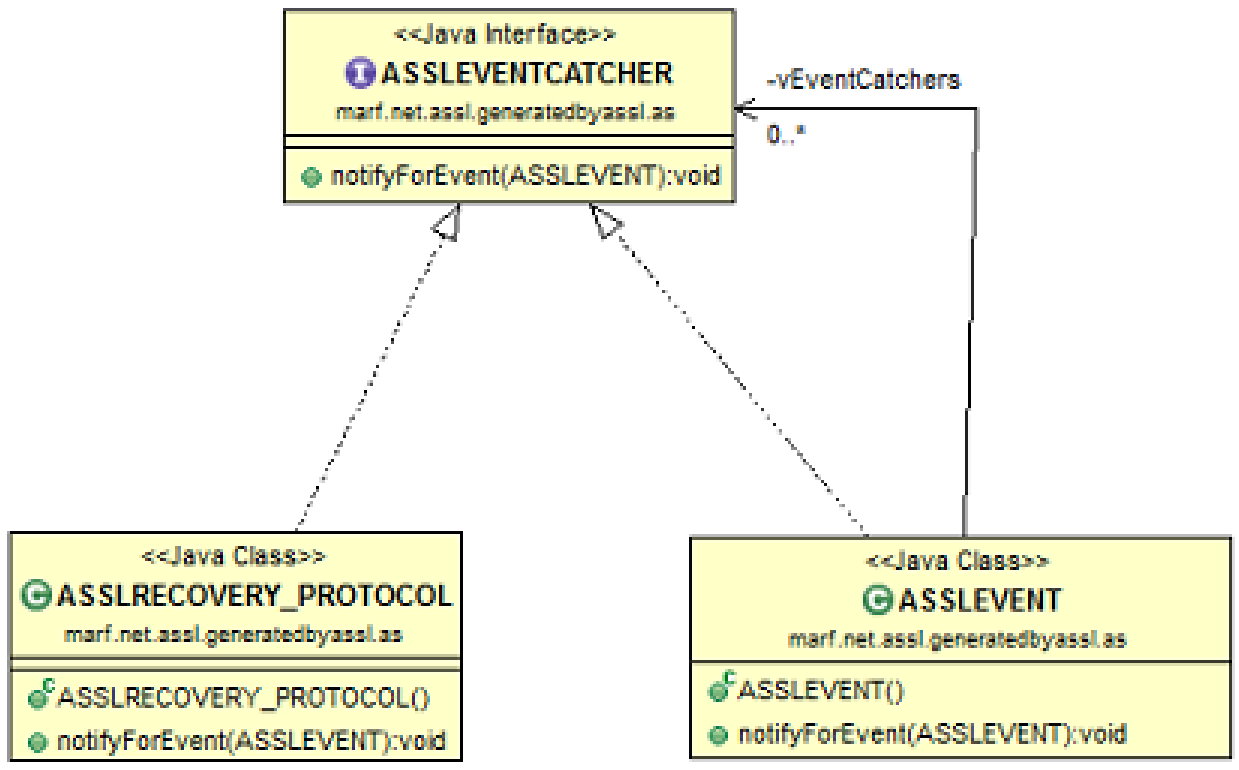

*Figure 26: Composite Pattern*

```
public interface ASSLEVENTCATCHER
{
public void notifyForEvent ( ASSLEVENT poEvent );
}
```

TABLE XVI. CODE SHOWING COMPONENT

```
public class ASSLEVENT extends Thread implements
ASSLEVENTCATCHER, ASSLMESSAGECATCHER
{....
public synchronized voidnotifyForEvent ( ASSLEVENT poEvent )
{
vOccurredEvents.add(poEvent);
}
....}
```

TABLE XVII. CODE SHOWING DECORATOR

```
public class ASSLRECOVERY_PROTOCOL implements
ASSLEVENTCATCHER
{....
```

```
public synchronized voidnotifyForEvent ( ASSLEVENT poEvent )
{
Enumeration<ASSLEVENT>eEVENTS = vInitiatedByEvents.elements();

ASSLEVENT currEvent = null;

while ( eEVENTS.hasMoreElements() )
{
currEvent = eEVENTS.nextElement();
if ( currEvent == poEvent )
{
save();
break;
}
}
}
}
```

TABLE XVIII. CODE SHOWING LEAF

### 4. PROTOTYPE

The classes participating to the Prototype Pattern are:
Client- creates a new object by asking a prototype to clone itself.
Prototype- declares an interface for cloning itself (if present).
Concrete Prototype - implements the operation for cloning itself. The process of cloning starts with an initialized and instantiated class. The Client asks for a new object of that type and sends the request to the Prototype class. A Concrete Prototype, depending of the type of object is needed, will handle the cloning through the clone() method, making a new instance of itself.

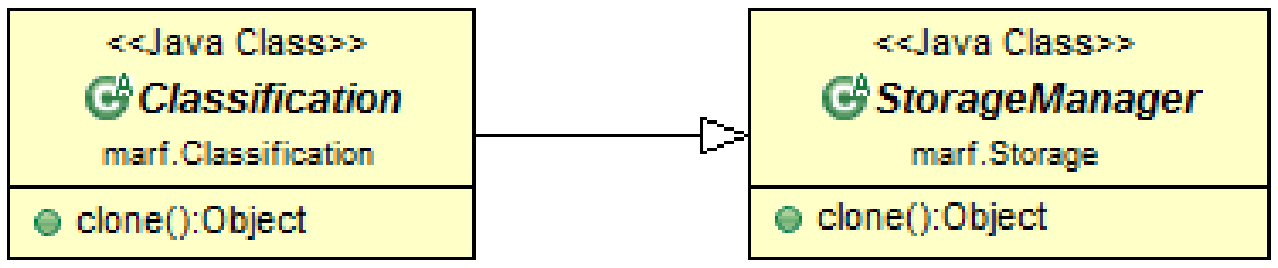

*Figure 27: PROTOTYPE Pattern*

Client: Classification
Prototype: StorageManager
Operation: clone()

- Client

```
public abstract class Classification
extends StorageManager
implements IClassification
{
public Object clone()
        {
                Classification oClone = (Classification)super.clone();
                oClone.oResultSet =
(ResultSet)this.oResultSet.clone();
                oClone.oTrainingSet =
(TrainingSet)this.oTrainingSet.clone();
                oClone.oFeatureExtraction = this.oFeatureExtraction;
                returnoClone;
        }
}
```

TABLE XIX. CODE SHOWING CLIENT

- Prototype

```
public synchronized Object clone()
{try
        {
        StorageManageroClone = (StorageManager)super.clone();

        oClone.strFilename = new String(this.strFilename);
        oClone.oObjectToSerialize = this.oObjectToSerialize;

        returnoClone;
                }

                // Should never happen.
        catch(CloneNotSupportedException e)
                {
                        thrownewInternalError(e.getMessage());
                }
        }
```

TABLE XX. CODE SHOWING PROTOTYPE

*b) GIPSY*

**1. Singleton pattern:**

Singleton pattern is a design pattern that restricts the instantiation of a class to
one object. This is useful when exactly one object is needed to coordinate actions across the system.

**Class Name:** DemandPool

```
Public class DemandPool
{
  private static DemandPoolsoInstance = null;

  public static synchronized DemandPoolgetInstance()
        {
                if(null == soInstance)
          {
                soInstance = newDemandPool();
          }

        returnsoInstance;
        }
}
```

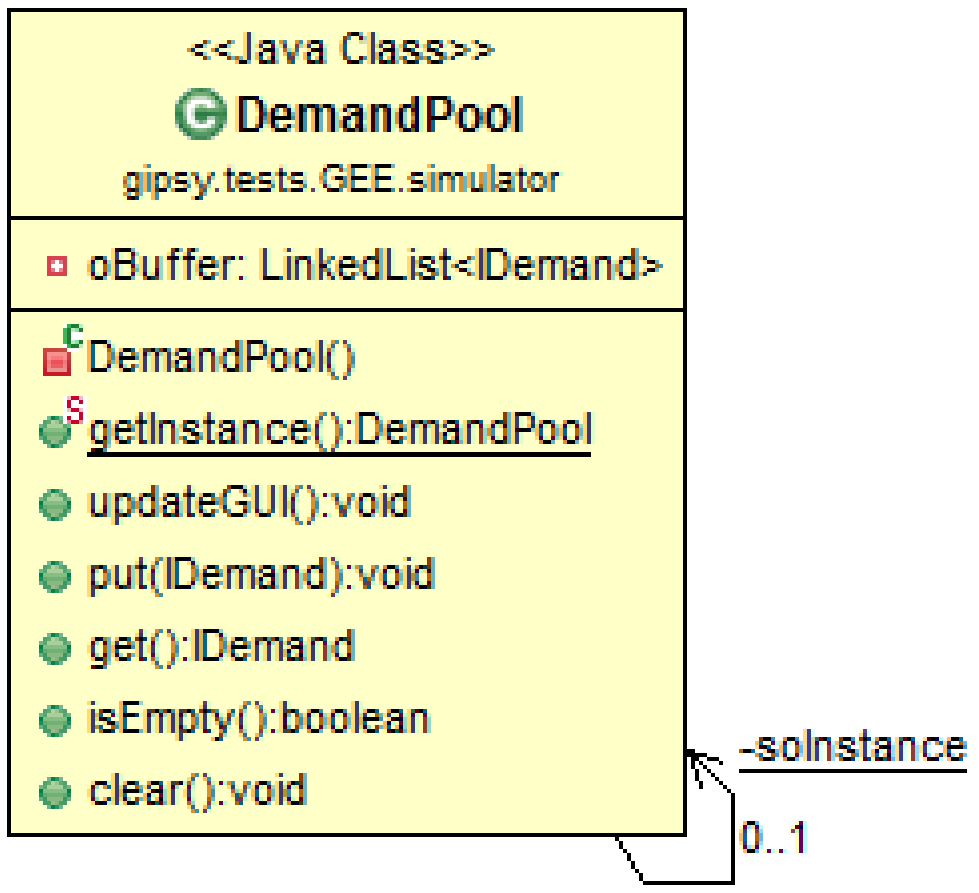

*Figure 28: SINGLETON Pattern of GIPSY*

**2. Observer Pattern:**

Observer pattern is a design pattern as per which, whenever there is one to many relationship between objects such as if one of the objects is modified, its dependent objects are to be notified automatically. This type of pattern falls under behavioural pattern category.Observer pattern uses three actor classes i.e. Subject, Observer and Client. Subject, an object having methods to attach and de-attach observers to a client object.

- **Observer**

```
public interface Idemand extends ISequentialThread, Cloneable
{............

DemandSignaturegetSignature();

 ..............}
```

TABLE XXI. CODE SHOWING OBSERVER PATTERN

- **Concrete Observer**

```
publicabstractclass
DemandextendsFreeVector<Object>implementsIDemand
{...........
DemandSignaturegetSignature();

publicDemandSignaturegetSignature()
        {
                returnthis.oSignature;
        }
...}
```

TABLE XXII. CODE SHOWING CONCRETE OBSERVER PATTERN

- **Subject**

```
public class ResultPool
{………

        publicsynchronizedvoidupdateGUI()
        {
                StringBuilderoStrBuilder = newStringBuilder();
                for(inti = 0; i<this.oBuffer.size(); ++i)
                {
oStrBuilder.append(this.oBuffer.get(i).getSignature().toString());
        oStrBuilder.append(GlobalDef.CR);
                oStrBuilder.append(GlobalDef.LF);
                }

        GlobalDef.soDGTDialog.getTAResults().setText(oStrBuilder.toStr
ing());
        }

…………}
```

TABLE XXIII. CODE SHOWING OBSERVER - SUBJECT

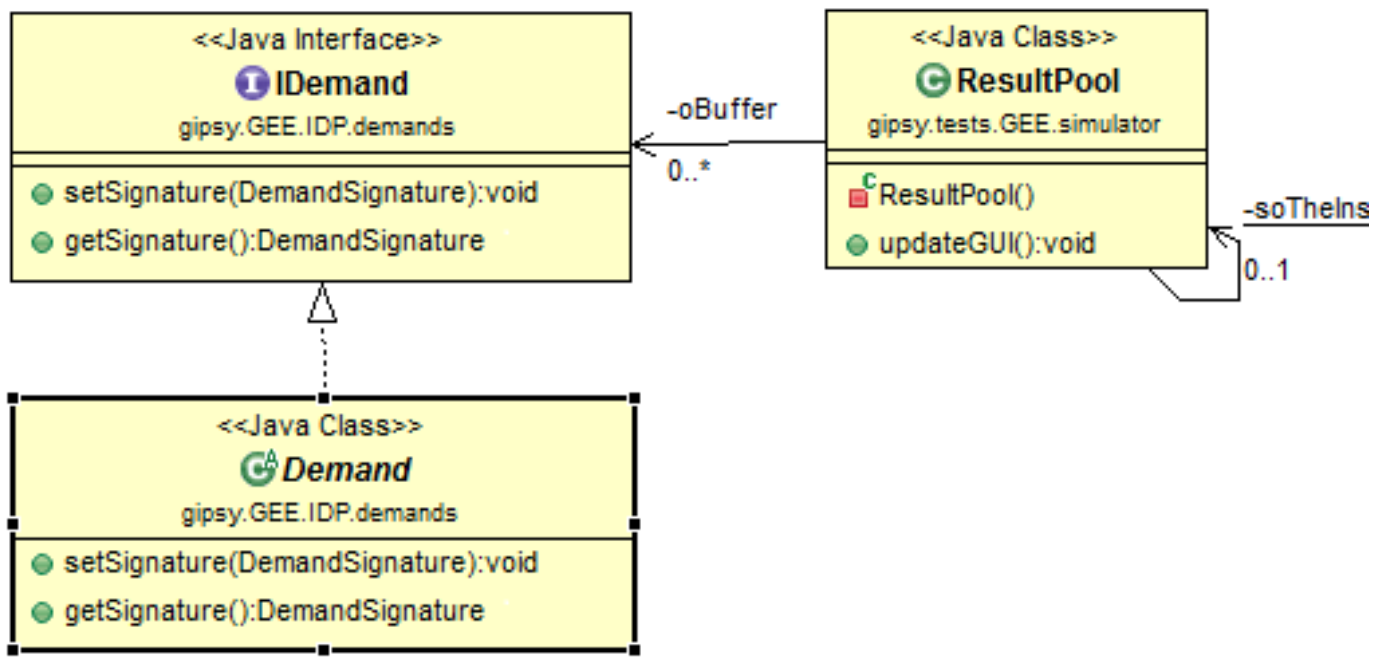

*Figure 29: Observer Pattern*

### 3. Factory Pattern:

In Factory pattern, objects are created without exposing the creation logic to the client and refer to newly created object using a common interface. This type of patterns fall under creational pattern and is one of the most used design patterns.

**Factory Method**

- **Creator**

```
public interface IdemandGenerator
{…….
    publicGIPSYTypeeval() throwsGEEException;
……}
```

TABLE XXIV. CODE SHOWING CREATOR

**Factory Method: publicGIPSYTypeeval()**

```
public abstract classDemandGeneratorextendsBaseThread
{…
        publicGIPSYTypeeval()
        throwsGEEException
        {
                returneval
                (
        (SimpleNode)this.oGEER.getAbstractSyntaxTrees()[0].getRoot(),
                        newGIPSYContext[] {
this.oGEER.getContextValue() },
                        0
                );
        }

…}
```

TABLE XXV. CODE SHOWING FACTORY METHOD

- **Product:**

```
GIPSYType
{
  …….
}
```

TABLE XXVI. CODE SHOWING PRODUCT

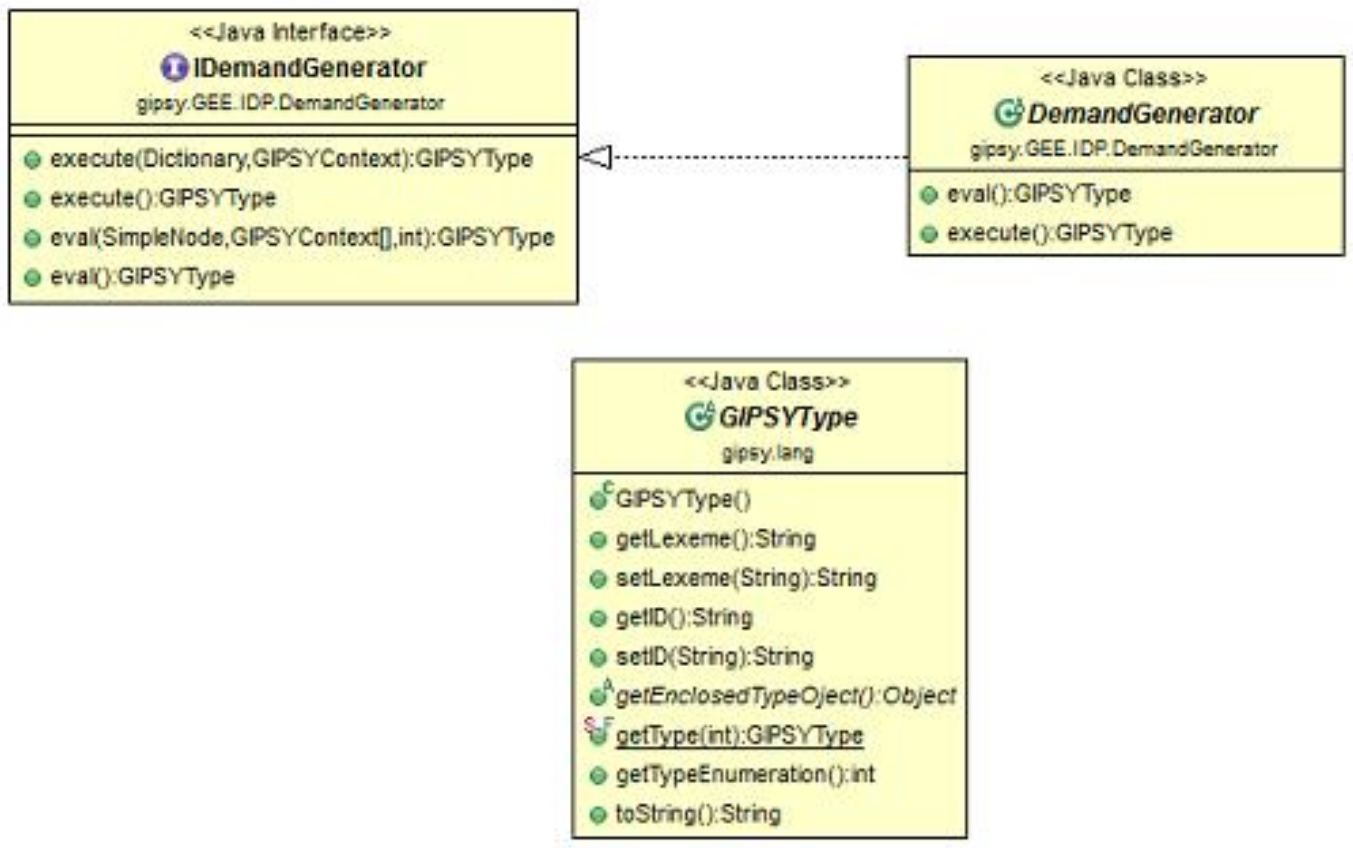

*Figure 30: Factory Pattern*

### 4. PROTOTYPE:

Prototype pattern refers to creating duplicate objects, while keeping performance in mind. This pattern falls under the category of creational pattern, as this pattern provides one of the best way to create an object. It involves a prototype interface which tells to create a clone of the current object. This is used when direct creation of objects is expensive.

**Client: GIPSYNode**

**Prototype: Configuration**

**Operation: run()**

**Client**

```
public class GIPSYNode extends Thread
{……..
public voidrun()
{
        Configuration oTierConfig = oRequest.getTierConfig();

oTierConfig = (Configuration) oRequest.getTierConfig().clone();
    }
…}
```

TABLE XXVII. CODE SHOWING CLIENT

**Prototype**

```
public class ConfigurationimplementsSerializable
{………

            public synchronized Object clone()
            {
            ConfigurationoNewConfig = newConfiguration();
            oNewConfig.setConfigurationSettings((Properties)
this.oConfigurationSettings.clone());
            returnoNewConfig;
            }
………}
```

TABLE XXVIII. CODE SHOWING PROTOTYPE

TABLE XXIX. SHOWS PATTERN NAMES AND CORRESPONDING MEMBER WHO WORKED ON RESPECTIVE PATTERN

| Pattern Names | Patterns found by: |
|---|---|
| GIPSY Patterns | |
| Singleton | Raveena Sharma |
| Prototype | Baljot Singh |
| Factory | Aman Ohri |
| Observer | Savpreet Kaur |
| | |
| DMARF Patterns | |
| Singleton | Navkaran Singh |
| Composite | Sukhveer Kaur |
| Prototype | Ravjeet Singh |
| Decorator | Manpreet Kaur |

## VI. Implementation

### a) God Class

Class Name: NeutralNetwork.java
We decided to refactor NeuralNetwork.java as the value of different metrics (WMC=
, ATFD=83, TCC=0.0) indicates that it is a god class.

The aim of the refactoring was to reduce the value of WMC by breaking this class into number of smaller classes. As the part of refactoring, we created two more classes.
The main logic was kept into the NeuralNetwork as the methods were dependent on each other and separating them would cause increase in coupling and low cohesion.

Class Name: NeuralNetworkParser.java
There was a static class named StorageManager which was parsing the XML files and was not dependent on the other methods, so we decided to move it to another class.

```
public class NeuralNetworkParser extends Classification implements
IStorageManager {
      protected NeuralNetworkParser(IFeatureExtraction
poFeatureExtraction)
      {
          super(poFeatureExtraction);
              }

              private StorageManager sm;

              public static class NeuralNetworkErrorHandler implements
ErrorHandler

      {……
              ……..}
}
```

TABLE XXX. REFACTORING CODE FOR GOD CLASS

Class Name: NeutralNetwork.java
We created a separate class for all the static variable.

```
public final class StaticVariables
{
public static final int DEFAULT_OUTPUT_NEURON_BITS = 32;
public static final double DEFAULT_TRAINING_CONSTANT = 1;
public static final String OUTPUT_ENCODING = "UTF-8";

………….}
```

TABLE XXXI. REFACTORING CODE FOR GOD CLASS – SEPARATE CLASS

After Refactoring, Metric Value: WMC=105, ATFD=87, TCC=0.0

### b) Duplicate Code

Class Name: Configuration_SOAPSerializer

We decided to refactor Configuration_SOAPSerializer.java as this class had repetition of same code in different if –else logics.
We created two methods setMember and setMemberState which contain the main logic. These methods were called by providing the different parameters as per the functionality required.

```
public void setMember(javax.xml.namespace.QName QName,
CombinedSerializer CombinedSerializer, XMLReader reader,
SOAPDeserializationContext context)
   {
    reader.nextElementContent();
    elementName = reader.getName();
    if (reader.getState() == XMLReader.START) {
       if (elementName.equals(QName)) {
          member = CombinedSerializer.deserialize(QName, reader, context);

          if(QName.equals(ns1_classificationMethod_QNAME) ){

instance.setClassificationMethod(((java.lang.Integer)member).intValue());
          }
          else if(QName.equals(ns1_currentSubject_QNAME))
          {

instance.setCurrentSubject(((java.lang.Integer)member).intValue());
          }
          else if(QName.equals(ns1_featureExtractionMethod_QNAME))
          {

instance.setFeatureExtractionMethod(((java.lang.Integer)member).intValue())
;
          }

          else if(QName.equals(ns1_preprocessingMethod_QNAME))
          {

instance.setPreprocessingMethod(((java.lang.Integer)member).intValue());
          }

          else if(QName.equals(ns1_sampleFormat_QNAME)){
          instance.setSampleFormat(((java.lang.Integer)member).intValue());
          }
          reader.nextElementContent();

       }
    }
   }



Configuration_SOAP.setMember(ns1_classificationMethod_QNAME,
ns3_myns3__int__int_Int_Serializer,reader, context);
Configuration_SOAP.setMember(ns1_currentSubject_QNAME,
ns3_myns3__int__int_Int_Serializer,reader, context);
```

TABLE XXXII. REFACTORING CODE FOR DUPLICATE CODE SMELL

#### c) *Long Method*

Class Name: GIPC
Long Method Name: process()
We decided to refactor this method as this this was very long and doing most of the comples tasks. We extracted two method from this by making sure that it is not impacting the the existing functionality.
New Methods: compileChunk() and groupAST()

```
public class GIPC extends IntensionalCompiler
{

public GIPSYProgram process()throws GIPCException{
        strPhase = compileChunk(oChunks,strPhase );
        strPhase = groupAST(strPhase) }
```

```
public String compileChunk(Vector<CodeSegment> oChunks, String
strPhase )
{

}

public String groupAST(String strPhase) throws GIPCException
{

}

}
```

TABLE XXXIII. REFACTORING CODE FOR LONG METHOD CODE SMELL

## VII. *Conclusion*

The study helped us understanding the functionality, implementation and properties of GIPSY and DMARF. Reverse engineering process was used to create the artifacts such as domain model, persona, and stakeholder identification, use case, design class diagram to understand the flow of code, stakeholders, work success scenario and relationship between classes. Various design patterns were identified and how they exist inside the code.
Different code smells that affected the code were recognized, what were they and why they were not good for the code. These smells were found using manual refactoring and, automatic refactoring tools such as Jdeodorant, PWD, JUnit, and CodePro. Refactoring was applied to tackle all the code smells. Refactoring helped towards betterment of the quality attributes by removing any code smells existing in the code.